\documentclass[manuscript,screen]{acmart}
\AtBeginDocument{%
  }

\setcopyright{none}
\copyrightyear{2025}
\acmYear{2025}
\acmDOI{XXXXXXX.XXXXXXX}
\acmConference[Conference acronym 'XX]{Pre-print}{2025}{Pre-print, AU}
\acmISBN{XXX-X-XXXX-XXXX-X/XXXX/XX}

\usepackage{xcolor}
\usepackage{framed}
\usepackage{stfloats}
\usepackage{algorithmic}
\usepackage{graphicx}
\usepackage{textcomp}
\usepackage{color}
\usepackage{pifont}
\usepackage{tabularray}
\usepackage{enumitem}
\usepackage{pgf-pie} 
\usepackage{pgfplots}
\pgfplotsset{compat=1.18} %
\usepackage{etoolbox} 
\usepackage{float}
\usepackage{rotating}
\usepackage{tikz}

\definecolor{formalshade}{rgb}{0.95,0.95,0.97}
\definecolor{darkblue}{rgb}{0.14,0.22,0.52}

\newcounter{takeaway}

\newenvironment{takeaway}{%
  \refstepcounter{takeaway}%
  \MakeFramed{\advance\hsize-\width\FrameRestore}%
  \noindent\textbf{Insight \thetakeaway: }\itshape
}{%
  \par\endMakeFramed\normalfont
}

\begin{document}

\title{Surveying GenAI-based Automation in Printed Circuit Board Design and Test}

\author{Sahana Srinivasan}
\email{sahana.srinivasan@unsw.edu.au}
\orcid{0009-0008-1304-7824}

\affiliation{%
  \institution{University of New South Wales}
  \city{Sydney}
  \state{New South Wales}
  \country{Australia}
}

\author{Benjamin Turnbull}
\affiliation{%
  \institution{University of New South Wales}
  \city{Canberra}
  \country{Australia}}
\email{benjamin.turnbull@unsw.edu.au}
\orcid{0000-0003-0440-5032}

\author{Hammond Pearce}
\affiliation{%
  \institution{University of New South Wales}
  \city{Sydney}
  \country{Australia}}
\email{hammond.pearce@unsw.edu.au}
\orcid{0000-0002-3488-7004}

\renewcommand{\shortauthors}{Srinivasan et al.}

\begin{abstract}
Generative artificial intelligence (GenAI) is increasingly used for applications in the hardware and software domains. 
It purports to reduce the manual effort involved in the development and testing of complex systems before release. 
Within the hardware space, most tasks have focused on design automation of integrated circuits, particularly with hardware description languages. 
However, other types of hardware also exist! 
In this survey, we instead examine how GenAI has been and is being across the printed circuit board (PCB) design life cycle. 
This includes everything from supply chains, system specification, circuit design, layout and optimisation, validation and test, and PCB assembly and distribution. 
Through this lens we present a taxonomy of discovered works, categorising them according to their intent and contributions. 
This survey also identifies key technical challenges that GenAI faces in this space, such as domain-specific data scarcity and limited support for integration with existing PCB tools. 
Finally, future research directions are discussed: our survey shows that there are many opportunities remaining when considering how GenAI may be integrated into various tasks in PCB design and test.
\end{abstract}



\maketitle

\section{Introduction}\label{sec:Intro}

\begin{figure*}[t]
    \centering
    \includegraphics[width=0.85\linewidth]{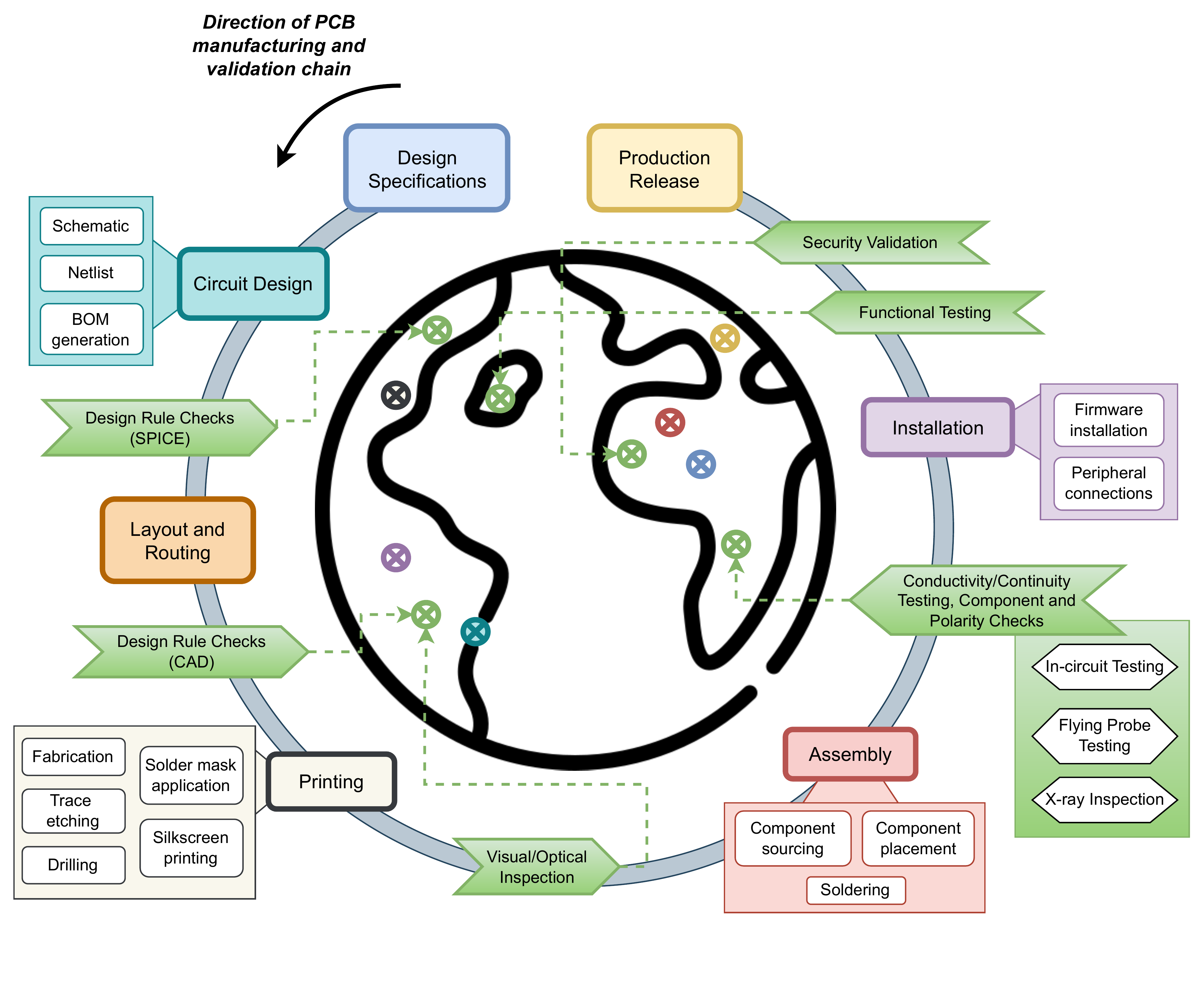}
    \caption{The simplified globalised PCB distributed supply chain model. For real companies, many parts of this process will be outsourced, and different softwares like SPICE or CAD enter at varying stages of design. Between each stage, there are tests performed that validate the previous stage, for example: physical testing, integration testing, etc. Research proposes the usage of GenAI for many tasks within this diagram. In this survey we explore the state of the art and identify open questions and areas which remain underexplored.}
    \label{fig:pcb-supply-chain-life-cycle}
\end{figure*}

Printed circuit boards (PCBs) are the backbone of all modern electronic devices \cite{asadizanjani_pcb_2017}. From smart watches to aerospace devices, medical devices and vehicles, modern devices feature computer systems built with PCBs.
The PCB design life cycle, as presented in Figure~\ref{fig:pcb-supply-chain-life-cycle}, follows a distributed supply chain model (sometimes termed a horizontal supply chain model \cite{asadizanjani_physical_2021}). This means many different groups and organisations are typically used in the tasks involved in producing a PCB, which include specification, architecture, integration and validation. Such distribution occurs because these tasks all require differing sets of tools, techniques, and increasingly specialised expertise.

PCBs begin their lives as sets of specifications and constraints which define the completed system. A bill of materials, or the components necessary for a functional system satisfying the specifications, is generated at this stage. Engineers produce and validate (e.g. through simulation) their schematics to check that those requirements are met \cite{chang_lamagic_2024}, before they are implemented in layout design files suitable for manufacturing. After the design rule checks and layout versus schematic tests are passed at both the schematic and layout generation stages respectively~\cite{you_interactive_2024}, they are sent to the physical manufacturer that prints the boards and runs their own tests (e.g. flying probes and inductive, etc.)~\cite{asadizanjani_physical_2021}. 

Once purchased electrical components listed in the bill of materials are placed and soldered to the PCB, automated optical inspections (e.g. X-ray~\cite{asadizanjani_pcb_2017, ghosh_advanced_2025} and other imaging methods~\cite{lin_design_2022}) validate successful assembly. 
The `bring-up' or installation of the PCB in the desired medium follows the assembly step, which might include additional steps such as firmware installation or physical connection to peripherals via additional wiring, if the intended application requires it. Since this stage is application-dependent, the time taken to perform functional tests varies from board to board. Finally, the resultant PCB is checked with integration tests to ensure the system is constructed as per initial requirements.

These complexities and the iterative validation model make PCB design and testing a difficult, time-consuming, and expensive process. 
Each vulnerability or flaw on the board also cascades throughout the life cycle \cite{asadizanjani_physical_2021}, which means that if it is not detected in the previous stage, it can increase the cost and manufacturing times before release \cite{singh_pcb_2024}.
Even when they are spotted early, these flaws can be difficult to patch and usually demand a reconstruction of the board: hardware is made up of physical artifacts not as amenable to post-implementation repair as in software systems.

Given these difficulties and the high cost of failure, it is desirable to introduce automation throughout the hardware design and product life cycle, and the field of Electronic Design Automation (EDA) has existed almost as long as electronics themselves \cite{wang_electronic_2009}. 
EDA seeks the reduction of tedious, manual, and error-prone processes. 
Examples include PCB schematic and layout software integrated with design rule checkers to ensure that properties like electrical conductivity and manufacturability remain sound \cite{xiang_digital_2024}. 

Recently, generative AI (GenAI) has made demonstrable advances in a huge variety of engineering domains, especially in tasks involving software code. 
Large Language Models are increasingly used to produce code \cite{chen_evaluating_2021}, debug code \cite{li_eda-debugger_2025}, explain code \cite{tafferner_can_2023}, educate new coders \cite{fruett_empowering_2024}, and more.
Hardware can also be considered like code, particularly with Integrated Circuits (ICs).
These often begin their implementation as files written in Hardware Description Languages (HDLs) such as Verilog and VHDL. 

As such, it is no surprise that LLMs and Vision Language Models (VLMs) such as \cite{radford_learning_2021}, which uses natural language to retrieve and caption new images, based on pre-trained labeled images, have shown promising results in reasoning and classification tasks. 
In the hardware manufacturing space, examples include \cite{zhou_causalkgpt_2024} which identifies aerospace manufacturing defects using a customized LLM, and in software domains, SQL-PaLM \cite{sun_sql-palm_2024} and CodeGeeX \cite{zheng_codegeex_2024} convert natural language instructions to code.

However, in the world of PCBs, research involving GenAI has been far more scant.
Though a small number of commercial options are beginning to appear, such as Flux.ai \cite{noauthor_better_2025} and Cadence OrCAD X \cite{cadence_home_pcb_2025}, there has been no concerted and widespread academic effort to comprehensively study the usage of GenAI in tasks for printed circuit boards. 

This work therefore sets out to understand the current state of the art of GenAI usage in the PCB lifecycle as a foundation for understanding and proposing future research directions in the field. This paper also identifies current limitations in the field and outlines research opportunities. 
To the best of our knowledge, this is the \textbf{first} survey to explore this area.

\subsection{Scope}

Our foundational survey sets out to determine the state of the art in applying GenAI within the broad domains of PCB Design and Test. 
Specifically, we seek to answer the following research questions:

\begin{itemize}[leftmargin=25pt]
    \item[\textbf{RQ1}] How is GenAI currently used to automate and improve PCB design, optimisation, validation and/or security verification processes, and how are researchers justifying any claims?
    \item[\textbf{RQ2}] What are the main technical challenges and limitations encountered when applying GenAI to PCB-related tasks, and how have recent studies addressed these obstacles?
    \item[\textbf{RQ3}] What future directions and opportunities exist for GenAI within PCB design and test to improve efficiency, accuracy, and/or scalability?
\end{itemize}

\subsection{Contributions}

As outlined in Section~\ref{sec:scope:keywords} and \ref{sec:scope:selection}, this work surveys 227 academic and industry papers adjacent to the research area, of which we include in this study 80 papers we can link to the application of GenAI in PCBs.
In Section~\ref{sec:applying-taxonomy}, the identified research is classified into a taxonomy linking GenAI utilisation to segments of the PCB design and test life cycle (see Figure~\ref{fig:pcb-supply-chain-life-cycle}).
We examine both what types of tasks GenAI are used for: be it design, optimisation or validation, 
as well as the various GenAI-based applications used in PCB contexts.
In Section~\ref{sec:discussions}, we analyse and discuss trends from the collected related work, compare them against the works we have taxonomised and draw comparisons and contrasts in our evaluations.
Section~\ref{sec:challenges-for-genai-usage} outlines what challenges and obstacles are present in the current literature and surveyed works with respect to GenAI applications in PCB design and test life cycle. 
Finally, through our taxonomy and analysed trends, we discover what the current gaps in academic literature are, and discuss future avenues in Section~\ref{sec:future-directions}.

\section{Background} \label{sec:Background}

In this section, we discuss the foundational concepts of PCB design and test and GenAI, and current contributions in the closely related domains such as hardware systems and other automation techniques.

\subsection{PCB Design}
Until the 1980s, PCBs were designed and manufactured manually with the help of stencils. 
As PCBs became more complex and were deployed across a wide range of electronic devices, from calculators to embedded systems and computers, the manual effort and repetitive nature of the design process became a significant bottleneck. 
These challenges motivated the development of electronic design automation (EDA) tools, to automate electronic design workflows and to enable scalable circuit and layout implementations~\cite{brayton_design_2015}. 

Alongside EDA's evolution, the modern PCB design and testing life cycle became a highly globalised and iterative process.
This is depicted in Figure~\ref{fig:pcb-supply-chain-life-cycle}. 
As each stage of the life cycle has become increasingly specialised, particularly in manufacturing and test, different organisations and teams have become increasingly capable at the tasks within, producing new tools and software to aid in those goals. As a life cycle, each stage of the production chain also depends on the timely competition of those before.
To coordinate the different steps in production, EDA tools can help.
PCB design software, such as the commercial Altium Designer \cite{noauthor_altium_2025}, or the open-source KiCad \cite{noauthor_kicad_2025} and freePCB \cite{noauthor_freepcb_2025}, are widely to support schematic generation, layout and routing, component placement, and design rule checks (to name but a few tasks). 
Nonetheless, all of these tools still remain reliant on human labour and expertise for creating and refining PCB designs.

\subsection{PCB Test}
Due to this completed and distributed design and testing life cycle for PCBs, mistakes -- including those which lead to security vulnerabilities -- can occur~\cite{alawandi_pcb_2025,harrison_malicious_2021}. For example: technical faults such as incorrect calibration during the solder deposition stage might result in an unintentional design flaw that causes thermal damage to security relevant components \cite{alawandi_pcb_2025}. 
These problems are further complicated by the lack of enforced standards for how electronic components are validated or functionally tested \cite{farzana_soc_2019}. 
Different businesses might focus on protecting or validating different parts of the board while unintentionally overlooking others. 
While an integrated validation approach covering a whole product's lifecycle is the most ideal strategy (e.g. Figure~\ref{fig:pcb-supply-chain-life-cycle} shows various intermediate testing stages embedded into the life cycle), actually implementing and performing such tests regularly can be difficult, and bugs continue to slip through to production. 
Recent notable examples include the 2025 recall of Medtronic Newport HT70/HT70 Plus Ventilators due to failing capacitors on the product's PCB~\cite{fda_medtronic_2025} and Ford's recall of certain Expedition/Navigator vehicles due to a battery junction box PCB which could develop an electrical short capable of overheating the PCB~\cite{dot_part_2025}.

However, not all defects in hardware designs are purely accidental. 
Hardware Trojans are malicious changes deliberately made to reduce product quality or introduce design `back doors' or triggers which may be later exploited to degrade performance or reduce privacy~\cite{rajendran_towards_2010, karri_trustworthy_2010}. Though much research has focused on chip-level hardware Trojans~\cite{adee_hunt_2008,xiao_hardware_2016}, PCB-level hardware Trojans have also been explored~\cite{paley_active_2016,mcguire_pcb_2019,piliposyan_hardware_2020,pearce_detecting_2022,krishnamurthy_muddle_2024} (particularly after the alleged hack on SuperMicro motherboards~\cite{mehta_big_2020,robertson_long_2021}).

One type of malicious attack induces aging effects in PCBs, and is called electro-migration based attacks \cite{mcguire_pcb_2019}. Changes to the cross-sectional areas of traces can make the associated nets more vulnerable to aging, but the effective change in trace resistance could be undetectable upon PCB inspection. Such changes could lead to broken circuits, or thermal damage of certain security-related components \cite{alawandi_pcb_2025}.

Given the twin challenges of functional and security bugs, and particularly given that such bugs and vulnerabilities may be added deliberately, there is a motivation for sound functional and security testing for PCBs throughout their design and product life cycles.

\begin{table*}[bp]
    \centering
    \definecolor{Concrete}{rgb}{0.949,0.949,0.949}
    \caption{Surveys of LLM applications in electronic design automation (EDA). Although these surveys claim to cover all of EDA, in actuality they are skewed towards ICs and do not examine PCBs.}
    \resizebox{0.90\textwidth}{!}{
    \begin{tblr}{
  row{even} = {Concrete},
  vline{2-3,6} = {-}{},
  hline{1,8} = {-}{0.08em},
  hline{2} = {-}{},
}
Source & Main focus & Design & Optimisation & Validation & Security\\
\cite{rai_enhancing_2025} & {Presents a \textit{taxonomy} of LLM applications in EDA for ICs} & \ding{51} & \ding{51} & \ding{51} & \\
\cite{he_large_2024} & How LLMs improve EDA, without examining PCBs & \ding{51} & \ding{51} & \ding{51} & \\
\cite{kande_llms_2024} & {Explores security risks and
vulnerabilities introduced\\by LLMs in hardware design and verification, but do not\\examine PCBs} & \ding{51} &  & \ding{51} & \ding{51}\\
\cite{qayyum_llms_2024} & {LLMs in hardware verification within EDA, without\\examining PCBs} &  &  & \ding{51} & \\
\cite{pan_survey_2025} & {Investigates LLMs usage in hardware architectures and\\the EDA workflow for ICs but does not evaluate PCB-\\based EDA} & \ding{51} &  &  & \\
\cite{xu_llm-aided_2024} & {LLM applications in HDL generation, \\debugging, verification, and implementation\\ while listing some
security concerns involved} & \ding{51} &  & \ding{51} & \ding{51}\\
\end{tblr}

    }
    \label{tab:llm_eda_survey}
\end{table*}

\subsubsection{Functional Testing}
There are certain standardized methods to functionally test or validate PCBs. It starts with design rule checks at the circuit schematic or netlist level \cite{xiang_digital_2024}, simulations to ensure the proposed design is valid \cite{kacmarcik_circuit_2022}, layout versus schematic tests \cite{you_interactive_2024} after the layout, placement or routing stage, and optical inspection at the least in the post-manufacturing stages.  

Given that correct component placement, orientation, and connectivity of all chips and associated nets on a PCB impact its functionality \cite{chen_pcbagent_2025}, functional testing also takes into account the product's performance after manufacturing.
This can begin as soon as the bare board is produced, e.g. with a flying probe test to check net connectivity, and/or can include similar tests once assembly is completed and components are soldered (i.e. PCBA). Further integration tests can occur after installation, depending on the intended application. 

During these tests, a PCB will be compared against their initial design specifications, with inputs necessary for the application simulated or emulated for the system \cite{aravind_balaji_development_2021}. 
Along with monitoring the behaviour of the PCB, electrical outputs (such as voltage drops), thermal differences and other physical measurements are observed for any abnormalities. This defect detection method is called side channel analysis. 
However, these strategies usually require human oversight, particularly when designing the test plans, which can become challenging for scalability reasons -- and, of course, noting that humans can also make mistakes when designing tests.

Once tests are designed, they can be automated. For example, computer vision-based approaches are automatic optical inspection methods for PCBs which can assist with detecting surface level flaws. Images of the board are acquired and preprocessed, from which components are isolated with feature extraction methods. These strategies then can identify components \cite{zhao_pcb_2022} and their misalignments, soldering problems and trace anomalies \cite{zhu_incorporating_2024}.  
There are also overall board level defect detection algorithms like \cite{sheng_visual_2024}, which is a machine learning based method and uses a lightweight VGG model. \cite{volkau_impact_2022} also is a deep learning model that extracts semantic patterns from defect-free samples and predicts flaws on PCBs.

\subsubsection{Security Validation}
Security validation requires PCB tests which focus not just on design functionality but instead on ensuring that a product meets its security requirements with respect to a given threat model. 
As an example of a recent security failure outside the scope of normal functional analysis, consider CVE-2025-26408~\cite{sec_consult_vulnerability_lab_cve-2025-26408_2025}, which saw an exposed JTAG become a vector for device takeover. From a functionality point of view, having such a debug port was not an issue; but given a motivated threat actor became a serious security vulnerability.

To perform security analysis requires expertise in a range of domains, including in identifying weak design patterns (e.g. Hardware CWEs \cite{noauthor_cwe_2025}), bugs which can become vulnerabilities, as well as determine areas of interest (i.e. security assets) which could become targets for malicious actors. Where designs are out-sourced, detecting malicious inclusions is also necessary.

One of the first approaches to an integrated security development life cycle was proposed by Microsoft \cite{howard_security_2006} in 2004.  
\cite{otieno_theory_2023} later suggested integrating security within the development process as an essential systematic design approach, that can ensure risks and security vulnerabilities are proactively fixed within every stage of the software development life cycle. 
Much of these principles apply even in the context of PCBs, as the design and test life cycle is an iterative model similar to software design.  
Given this security development life cycle, it can be seen that coordination between hardware design engineers and security professionals is essential for a functionally operational and secure system. Both teams must understand what components and associated nets could potentially offer unauthorised access or leak data. Similarly, mitigation strategies must be discussed and agreed upon.

At present, no automated tool has all the relevant context or the domain knowledge of a security professional~\cite{ghimire_hardware_2025}, which leaves much of security-based automation under-researched.
However, manual security validation methods are also prone to human error or inconsistencies, which could cascade through the supply chain and cause delays. 
Just like the iterative bug-fixing approach for PCB design, an integrated security validation approach \cite{asadizanjani_physical_2021}, where vulnerabilities are scanned for in each design validation step, is desirable for testing in the PCB life cycle. 

\subsubsection{Current Advances with GenAI for PCB Design and Testing}
Although EDA tools automate some of the PCB design and testing stages, the life cycle in Figure~\ref{fig:pcb-supply-chain-life-cycle} relies heavily on human intervention, particularly for making informed decisions, defining constraints, and iteratively validating designs. 
Recent advances in GenAI address some of these limitations with the help of LLMs.
Table~\ref{tab:llm_eda_survey} surveys some recent GenAI works in the EDA domain, and shows that significant progress has been made at the integrated circuit (IC) level, and in broader electronic design contexts as well. 
For instance, \cite{pan_survey_2025} and \cite{rai_enhancing_2025} highlight LLM-enabled enhancements to EDA workflows, both of which focus primarily on ICs. 
Similarly, \cite{he_large_2024}, \cite{kande_llms_2024}, and \cite{qayyum_llms_2024} investigate LLM applications in electronics design, but not specifically at the PCB level. Despite this, the rising trends in EDA and GenAI usages in electronic spaces suggest board level automation with the help of LLMs, or other GenAI avenues, to be a promising area of research.

Given this, in Section~\ref{sec:RelatedWork}, we explore GenAI applications in electronic design and testing, and set the context for our new taxonomy for GenAI-based PCB automation in Section~\ref{sec:applying-taxonomy}.

\section{Related Work} \label{sec:RelatedWork}
Though GenAI in the form of LLMs is being applied to a variety of manufacturing tasks and contexts~\cite{zhao_special_2025,khan_chatgpt_2025}, in this survey, we are most interested in the design and manufacturing of PCBs, and this section presents an overview of how current GenAI is intersecting this area. Table~\ref{tab:relatedWorkcomparison} outlines some related work in the domains of embedded and cyber-physical systems, IC or chip-based design, analog circuit design, and applied educational tools for electronic design that GenAI models are used in. 

Noting that there is currently no survey specifically in GenAI based applications for PCBs, we critically analyse new techniques, processes and shed some light on some identified research opportunities from the overarching field. We aim to use this knowledge and the insights we obtain as a foundation for our survey.

\begin{table*}[tp]
    \centering
    \definecolor{Concrete}{rgb}{0.949,0.949,0.949}
    \caption{GenAI applications in embedded systems and cyber-physical systems, chip-based architectures, analog circuit domains, and applied domains such as education of electronic concepts and tools. This table highlights certain trends in the neighboring domains that can prove insightful for what is expected in PCBs as well.} 
    \resizebox{0.95\textwidth}{!}{\begin{tblr}{
  row{even} = {Concrete},
  column{1} = {Concrete},
  cell{2}{1} = {r=7}{c},
  cell{9}{1} = {r=2}{c},
  cell{11}{1} = {r=4}{c},
  cell{15}{1} = {r=4}{c},
  vline{2-4,7} = {-}{0.08em},
  hline{1,19} = {-}{0.08em},
  hline{2,9,11,15} = {-}{},
}
{Related\\Work\\Category} & Source & RQ? & Design & Optimisation & Validation & Security\\
{\begin{sideways}{\shortstack{Embedded systems or cyber-~\\physical systems}}\end{sideways}} & \cite{englhardt_exploring_2024} & {LLMs for debugging support in embedded system \textit{programming} with\\a physical testbench of sensor-actuator pairs} & \ding{51} &  & \ding{51} & \\
 & \cite{sabree_openai_2024} & {ChatGPT for embedded system \textit{programming}} & \ding{51} &  &  & \\
 & \cite{jansen_words_2023} & {Benchmarks ChatGPT-4 and Claude-V1 against MICRO25 and PINS100 for\\simple electronic device design tasks (like turning on an LED)} & \ding{51} &  & \ding{51} & \\
 & \cite{xu_llm-enabled_2024} & {Categorises existing LLM-enabled CPS applications, and lists their\\security concerns} & \ding{51} &  &  & \ding{51}\\
 & \cite{lu_enabling_2024} & {Presents a design tool enhanced by LLMs for embedded mechanical \\computation design ideation} & \ding{51} &  &  & \\
 & \cite{cui_large_2024} & A multi-agent LLM-based framework to automate controller design & \ding{51} & \ding{51} & \ding{51} & \\
 & \cite{sarhaddi_llms_2025} & {Illustrates data processing with LLMs in Internet of Things (IoT) systems, and\\lists their security threats} & \ding{51} & \ding{51} &  & \ding{51}\\
\begin{sideways}{\shortstack{IC/Chip~\\Design}}\end{sideways} & \cite{mirka_gannoc_2021} & {GAN-based framework that automatically generates customised \\Network-on-Chip topologies to improve the classification model} & \ding{51} & \ding{51} &  & \\
 & \cite{kashyap_generative_2023} & {Modifying the power and thermal properties of a chip’s design using\\conditional GANs} &  & \ding{51} &  & \\
\begin{sideways}\shortstack{Analog~\\Circuit }\end{sideways} & \cite{guo_circuit_2019} & {A GAN-based framework for circuit synthesis that overcomes\\traditional circuit enumeration limitations, enables scalable circuit\\topology generation and offers deeper design exploration.} & \ding{51} & \ding{51} & \ding{51} & \\
 & \cite{he_generative_2020} & {A GAN and cross-wavelet-transform-based deep learning method for analog\\circuit fault diagnosis that effectively captures time–frequency features and\\improves classification with limited training data.} &  &  & \ding{51} & \\
 & \cite{gui_fault_2025} & {Feedforward and recurrent network based approach, with an\\adjustable weighting scheme, to achieve accurate and reliable analog-\\circuit fault diagnosis across diverse circuit types and environments.} &  &  & \ding{51} & \\
 & \cite{ho_denoising_2020} & {An image based framework that de-noises circuit images with diffusion\\models and a strategy called auto-regressive decoding.} & \ding{51} & \ding{51} &  & \\
\begin{sideways}\shortstack{Applied educ-\\ation tools}\end{sideways} & \cite{fruett_empowering_2024} & {An interactive open source hardware platform that acts as a tutor and integrates\\hardware and software knowledge in Science, Technology, Engineering, Arts, and\\Mathematics domains} & \ding{51} & \ding{51} & \ding{51} & \\
 & \cite{elder_can_2023} & {Can AI be used to solve digital design problems given a lab sheet of instructions?} & \ding{51} &  & \ding{51} & \\
 & \cite{meshram_electrovizqa_2024} & {Can Multi-modal LLMs effectively understand and solve visual digital electronics\\problems, and how can a proposed domain-specific benchmark evaluate and\\measure performance?} & \ding{51} &  & \ding{51} & \\
 & \cite{li_eee-bench_2025} & {How do Large Multi-modal Models perform when fed with real-life engineering\\problems? This benchmark is tested against electrical and electronics engineering\\(EEE) questions.} & \ding{51} &  & \ding{51} & \\
\end{tblr}
}
    \label{tab:relatedWorkcomparison}
\end{table*}

\subsection{Electronics Design and Testing} \label{sec:RelatedWork:elecDesignAndTest}

PCBs constitute a specialised subset of electronic design that focuses on translating circuit schematics into manufacturable physical layouts, which are then validated to produce fully functional boards. The broader field, however, encompasses several other hardware architectures and components. For these reasons, in this subsection, we explore GenAI works in bigger hardware architectures like embedded systems and cyber-physical systems, and other electronic and electrical components comprised in them. Finally, we make note of domains closely related to electronics and how GenAI is used in that space to analyse rising trends. All these works are tabulated in Table~\ref{tab:relatedWorkcomparison} and analysed based on whether they are design modifications, optimisation improvements or validation based: either functional or security relevant.

\textbf{Embedded systems and cyber-physical systems} are electronic systems that are designed to perform or optimise a specific real-world task \cite{xu_embedagent_2025}. Design and test here includes micro-controller or processor selection, peripheral integration, firmware development, and timing verification, as well as the incorporation of sensors, actuators, and networked control for real-world interactions. Testing in these domains is multifaceted, including functional verification, real-time performance evaluation, reliability assessment, and fault diagnosis. 

As shown in Table~\ref{tab:relatedWorkcomparison}, GenAI-based works in this area include \cite{englhardt_exploring_2024} and \cite{jansen_words_2023}. %
While \cite{englhardt_exploring_2024} uses a self-proposed automated test-bench, \cite{jansen_words_2023} analyses how two LLMs compare against one another in electronic device design using its two proposed datasets. No security risks are listed in these works. Meanwhile, \cite{xu_llm-enabled_2024} and \cite{sarhaddi_llms_2025} complete security risk assessments in their works but neither of these are methodology based. They do not create automated strategies for security validation to bridge existing research gaps. As a result, for embedded systems and cyber-physical systems, Table~\ref{tab:relatedWorkcomparison} highlights significant research opportunities at the intersection of security threats and functional validation spaces.

\textbf{Network-on-Chip} (NoC) is a communication architecture enabling efficient data transfer among on-chip blocks. \cite{mirka_gannoc_2021} is one such framework that uses \textit{Generative Adversarial Networks}, a type of deep-learning technique that generates and augments data \cite{gonog_review_2019} using image-based inputs. Because of their ability to generate synthetic data in a single step, in complex systems such as NoCs, GANs can generate synthetic designs that satisfy certain desired properties and achieves improved connectivity compared to the training dataset statistics. On the other hand, at the 
\textbf{chip level}, 
GANs is useful for electronic component defect detection \cite{zhang_defect-gan_2021}, where diverse synthetic images of rare defects can be generated to improve training of inspection models. 
\cite{kashyap_generative_2023} specifically focuses on optimisation of the power and thermal properties of a chip's design using conditional GANs. These classification tasks help in determining design flaws, that could reveal functional and/or security vulnerabilities as well.

\textbf{Analog circuits} also play a vital role in electronic boards, as without power sources, resistors or capacitors, the availability of a hardware system is impacted. For tasks such as denoising, \textit{Diffusion Models} has gained significant traction as their outputs are high-quality images, audio, and other data types useful for augmented or extended datasets \cite{ho_denoising_2020}.
Similarly, for circuit synthesis, GANs can be used to generate and optimise analog circuit topologies \cite{guo_circuit_2019}, which can solve larger-scale circuit synthesis problems and provide deeper design insights than traditional enumeration methods. 
These design generation methods overlap closely with machine-learning based algorithms frameworks such as \cite{he_generative_2020} and \cite{gui_fault_2025}, both of which perform fault detection and diagnosis using neural networks. The line between ML and GenAI blurs because of the classification-based problems that could fall into either buckets. Since ML algorithms are a rabbit-hole of their own, in this survey, we do not explore these recent works in detail.

\textbf{Applied educational tools for the field of electronic design: }
At a more applied scale, LLMs have also been applied to electronic design teaching and education. Specifically, \cite{elder_can_2023} tested ChatGPT's ability to complete a sophomore-level digital design laboratory task and found that though it was ill-equipped to determine wiring and routing of circuits, ChatGPT was able to answer lab questions, ultimately receiving a grade of 73\% for the task assessment. \cite{meshram_electrovizqa_2024} and \cite{li_eee-bench_2025} are tools specifically useful to understand and solve digital electronic circuit questions or problems; the former is an LLM model while the latter is a Large Multimodal Model (LMM). BitDogLab \cite{fruett_empowering_2024} is another education-oriented tool designed to promote learning in embedded systems, programming, and electronics. 
These works illustrate that LLMs have the potential to communicate and integrate with existing EDA frameworks to solve open-ended problems. However, current EDA tools do not allow easy integration of LLMs, which limits GenAI research opportunities in this area. EDA improvements that diverge from PCB usages are excluded from our survey.

\subsection{Summary and Research Opportunities}
GenAI, especially LLMs, have been predominantly used to improve engineering automation, enhance electronic design and test life cycles and show active research in the design and optimisation stages. There is no existing survey, however, that lists or critically analyses their contributions against one another. Therefore, our survey is a first of its kind in the PCB and GenAI space, that highlights current trends, known limitations or knowledge gaps, and areas for future work.

From Section~\ref{sec:RelatedWork:elecDesignAndTest}, we note that GenAI is used for very specific applications in the electronics domain. Broadly, these works can be classified into three divisions: Design (which includes improvements and modifications to the electronic design, and involves tinkering with EDA tools to change schematics or layouts), Optimisation (which can be either performance-based improvements, spatially compact design modifications, and/or power or thermal-based improvements), and Validation (under which defect detection, functional testing, and security verification fall under). 

In the context of GenAI applications to PCB design, improvements to the design and testing life cycle is still nascent. There are multiple related works that progress the field in unique ways, but the field is still missing a cohesive understanding of the different drivers and related areas. These are taxonomised in Section~\ref{sec:applying-taxonomy} to compare and contrast research in the space. 

\section{Survey Methodology and Scope}
\label{sec:scope}

We set out this survey to identify the various applications of GenAI in PCBs following the RQs outlined in Section~\ref{sec:Intro}. We begin by searching keywords outlined in Section~\ref{sec:scope:keywords} and define inclusion and exclusion criteria according to Section~\ref{sec:scope:selection}. 
From this, we have surveyed 227 papers and included a total of 80 papers, as of August 8, 2025.
The collected papers are categorised based on the PCB design and test life cycle stage at which the selected work can be applied. 
We also determine whether it assists with design generation, fine-tuning or optimisation, or functional or security validation. These rules that form the basis of our taxonomy are represented as a table defined in Table~\ref{tab:taxonomy_table}. 
We finally present the findings of our survey based on each manufacturing stage of PCB design and test in Section~\ref{sec:applying-taxonomy}. 

\subsection{Searched Keywords}
\label{sec:scope:keywords}

\begin{table*}[tp]
    \centering
    \caption{Search term categorisation in the context of the GenAI-based PCB survey.}
    \resizebox{\textwidth}{!}{\definecolor{Concrete}{rgb}{0.949,0.949,0.949}
\begin{tblr}{
  row{even} = {Concrete},
  cell{3}{2} = {r=4}{},
  cell{7}{2} = {r=7}{},
  vline{2,3} = {-}{},
  hline{1,14} = {-}{0.08em},
  hline{2-3,7} = {-}{},
}
Term & Purpose of Search & Scope Search Phrase\\
{Generative AI\\(GenAI)} & {Umbrella term \\for works relevant\\to the survey} & {"Printed Circuit Boards and GenAI" OR "PCBs and GenAI" OR "Printed Circuit Boards and Generative AI" OR~\\"PCBs and Generative AI"}\\
{Large Language\\Models (LLMs)} & {GenAI based\\themes of \\PCB work} & "Printed~Circuit Boards and Large Language Models" OR "Printed Circuit Boards and LLMs" OR "PCBs and LLMs"\\
{Vision Language\\Models (VLMs)} &  & "Printed~Circuit Boards and Vision Language Models" OR "Printed~Circuit Boards and VLMs" OR "PCBs~and VLMs"\\
{Generative \\Adversarial \\Networks (GANs)} &  & {"Printed~Circuit Boards and Generative Adversarial Networks" OR "Printed~Circuit Boards and GANs" OR "PCBs and GANs"}\\
Diffusion Models &  & "Printed~Circuit Boards and Diffusion Models" OR "PCBs~and Diffusion Models"\\
Design Specification & {Generative \\(GenAI) works\\against each of\\the PCB design \\and test life \\cycle stages~\\ \\Note: Each search\\query was preceded\\by "PCB"} & {"Design Specification and Large Language Models" OR "Design Specification and LLMs"~OR~"Design Specification and\\Vision Language Models"~OR "Design Specification and VLMs" OR"Design~Specification and Generative Adversarial \\Networks"~OR~"Design Specification and GANs" OR "Design~Specification and Diffusion Models"}\\
Circuit Schematic &  & {"Circuit Schematic and Large Language Models"~OR "Circuit Schematic and LLMs"~OR~"Circuit Schematic and Vision \\Language Models"~OR~"Circuit Schematic and VLMs" OR "Circuit Schematic and Generative Adversarial Networks"~\\OR~"Circuit Schematic and GANs" OR "Circuit Schematic and Diffusion Models"}\\
Netlist Generation &  & {"Netlist Generation and Large Language Models"~OR "Netlist Generation and LLMs"~OR~"Netlist Generation and Vision\\Language Models"~OR~"Netlist Generation and VLMs" OR "Netlist Generation and Generative Adversarial Networks"~OR\\"Netlist Generation and GANs" OR "Netlist Generation and Diffusion Models"}\\
Layout or Routing &  & {"Layout or Routing and Large Language Models"~OR "Layout or Routing and LLMs"~OR~"Layout or Routing and Vision\\Language Models"~OR~"Layout or Routing and VLMs" OR "Layout or Routing and Generative Adversarial Networks"~OR~\\"Layout or Routing and GANs" OR "Layout or Routing and Diffusion Models"}\\
Printing &  & {"Printing and Large Language Models"~OR "Printing and LLMs"~OR~"Printing and Vision Language Models"~OR~"Printing\\and VLMs" OR "Printing and Generative Adversarial Networks"~OR~"Printing and GANs" OR "Printing and Diffusion Models"}\\
Assembly &  & {"Assembly and Large Language Models"~OR "Assembly and LLMs"~OR~"Assembly and Vision Language Models"~OR\\"Assembly and VLMs" OR "Assembly and Generative Adversarial Networks"~OR~"Assembly and GANs" OR "Assembly \\and Diffusion Models"}\\
Installation &  & {"Installation and Large Language Models"~OR "Installation and LLMs"~OR~"Installation and Vision Language Models"~OR\\"Installation and VLMs" OR "Installation and Generative Adversarial Networks"~OR~"Installation and GANs" OR "Installation\\and Diffusion Models"}
\end{tblr}
}
    \label{tab:searched_keywords}
\end{table*}

Relevant literature for our survey is collected using certain keywords directly relevant to our field. The search has been conducted iteratively, starting from a broad, umbrella term for works of GenAI in PCBs relevant to the survey. 

Once a few papers were collected on this topic, their abstracts and keywords sections were skim-read for general trends of GenAI, and which models were primarily used for automation. Four categories of foundational models were obtained, namely: LLMs, VLMs, GANs and diffusion models. The next stage of search comprised on model-specific queries, particular to printed circuit boards. The works obtained roughly fell into one of the PCB design and test life cycle stages described in Figure~\ref{fig:pcb-supply-chain-life-cycle}. 

We hence extended our search to stage-wise GenAI based applications and gathered all the papers published in these areas. Table~\ref{tab:searched_keywords} outlines the exact scope queries used in the collection of works. %

\subsection{Selection Criteria} 
\label{sec:scope:selection}

In order to identify the papers directly related to our survey of GenAI applications on PCBs, we formulated a set of inclusion and exclusion criteria. 

From the 227 collected papers, we manually reviewed and filtered the papers that did not contain any relevance to the keywords listed in Section~\ref{sec:scope:keywords}. For example: papers that used the term PCB but did not imply printed circuit boards were manually removed from the collection upon manually reading the abstract. 
Any papers that were closely linked to the PCB design and test life cycle, like analog circuit design or component design automation, were included for the purposes of our proposed taxonomy in Section~\ref{sec:applying-taxonomy}.

This process of screening papers through a predefined filter improves the transparency of our survey and allows for reproducibility of our study. Table~\ref{tab:inclusion_and_exclusion_criteria_tab} presents the inclusion and exclusion criteria used in this survey.

\begin{table}[h]
    \centering
    \caption{The scope criteria against which the collected papers are assessed.}
    \resizebox{0.80\textwidth}{!}{\definecolor{Concrete}{rgb}{0.949,0.949,0.949}
\begin{tblr}{
  row{5-7} = {Concrete},
  cell{2}{1} = {r=3}{},
  cell{5}{1} = {r=3}{},
  vline{2} = {1-7}{},
  hline{2,5} = {-}{},
  hline{3-4,6-7} = {2}{},
  hline{1,8} = {-}{0.08em},
}
Criteria & Conditions\\
Inclusion & {GenAI applications in PCB related tasks}\\
 & {Novel methods, tools, or experimental evaluations of GenAI (specifically focusing\\on GANs, diffusion models, LLMs and VLMs) in PCBs}\\
 & {Peer-reviewed journal articles, conference papers, or preprints published in English\\within the last 5 years}\\
Exclusion & {Specialised focus on chip design automation (eg. Verilog generation) rather than PCB level}\\
 & {GenAI usages outside PCB contexts (eg. in embedded systems or cyber-physical systems} \\
 & {Software-focused GenAI applications (eg. firmware implementation or installation)} \\
\end{tblr}
}
    \label{tab:inclusion_and_exclusion_criteria_tab}
\end{table}

\subsection{Quantitative Analysis}
\label{sec:scope:quantitative
}
In this subsection, we gather and analyse the general trends in our collected papers and make some initial observations based on numbers. 
We mainly explore what the distribution of papers looks like, in terms of year of publishing, whether or not they match the selection criteria detailed in Section~\ref{sec:scope:selection}, and how many papers directly link to the different PCB design and test stages from Figure~\ref{fig:pcb-supply-chain-life-cycle}. 

\subsubsection{Temporal Distribution of Collected Papers} 
As of the data that was collected by August 8, 2025, Figure~\ref{fig:5yearsOfPapersLLMs} graphically represents the distribution of the 227 collected articles published in the last five years, 2021 inclusive.

\begin{figure}[htbp]
  \centering
  \begin{minipage}{0.48\textwidth}
    \centering
    \resizebox{0.95\linewidth}{!}{\begin{tikzpicture}
  \begin{axis}[
    ybar,
    bar width=16pt,
    width=10cm,
    height=7.5cm,
    ylabel={Number of papers},
    xlabel={Year},
    ymin=0,
    ymax=100,
    symbolic x coords={2021,2022,2023,2024,2025},
    xtick={2021,2022,2023,2024,2025},
    nodes near coords,
    nodes near coords align={vertical},
    every node near coord/.append style={font=\small,color=black},
    every axis plot/.append style={
        ybar,
        line width=0.5pt,
        bar shift=0pt,
        bar width=16pt
    },
    enlarge x limits=0.15,
    ymajorgrids=true,
    grid style=dashed,
  ]
    \addplot+[draw=black,fill=red!25] coordinates {(2021,9)};
    \addplot+[draw=black,fill=orange!25] coordinates {(2022,15)};
    \addplot+[draw=black,fill=purple!25] coordinates {(2023,25)};
    \addplot+[draw=black,fill=green!25] coordinates {(2024,87)};
    \addplot+[draw=black,fill=cyan!25] coordinates {(2025,74)};
  \end{axis}
\end{tikzpicture}}
    \caption{Temporal distribution of the surveyed GenAI and PCB related works published between 2021 and 2025. It indicates a sudden shoot-up of research activity trends over the last two years. Data is accurate as of August 8, 2025.}
    \label{fig:5yearsOfPapersLLMs}
  \end{minipage}\hfill
  \begin{minipage}{0.48\textwidth}
    \centering
    \resizebox{0.95\linewidth}{!}{\begin{tikzpicture}
  \begin{axis}[
    xbar,
    bar width=12pt,
    width=10cm,
    height=7.75cm,
    xlabel={Number of Papers},
    ylabel={Database},
    symbolic y coords={ACM, IEEE, MDPI, SciDir, arXiv, Other},
    ytick={ACM, IEEE, MDPI, SciDir, arXiv, Other},
    y dir=reverse,
    ytick style={draw=none},
    nodes near coords,
    nodes near coords align={horizontal},
    every node near coord/.append style={font=\small,color=black},
    every axis plot/.append style={
        xbar,
        draw=black,
        line width=0.5pt,
        bar shift=0pt,
        bar width=12pt
    },
    enlarge y limits=0.2,
    xmajorgrids=true,
    grid style=dashed,
    xmin=0,
    xmax=100
  ]
    \addplot+[draw=black,fill=red!25] coordinates {(11,ACM)};
    \addplot+[draw=black,fill=cyan!25] coordinates {(72,IEEE)};
    \addplot+[draw=black,fill=orange!25] coordinates {(16,MDPI)};
    \addplot+[draw=black,fill=gray!25] coordinates {(24,SciDir)};
    \addplot+[draw=black,fill=purple!25] coordinates {(45,arXiv)};
    \addplot+[draw=black,fill=green!25] coordinates {(66,Other)};
  \end{axis}
\end{tikzpicture}}
    \caption{Distribution of the collected papers according to their identified source repositories. All Arxiv references are identified from Google Scholar. Data is accurate as of August 8, 2025.}
    \label{fig:barChartDigitalLibraryDistribution}
  \end{minipage}
\end{figure}

We note a general trend of increasing number of publications in the intersection of GenAI-based applications and PCBs, which is consistent with our understanding of generative AI, its growing capabilities and rapid advancements in the hardware and software domains. 

\subsubsection{Searched Databases of Collected Papers} 
The digital databases papers were collected from are: ACM, arXiv, IEEE Xplore, MDPI, ScienceDirect and other resource hubs. 
If the source could not be identified as a publication in ACM, IEEE Xplore, MDPI or ScienceDirect, we examine if it is solely a preprint on arXiv and mark it if so, otherwise we choose Other. 
Figure~\ref{fig:barChartDigitalLibraryDistribution} illustrates the distribution by the online database the source was obtained from.

From Figure~\ref{fig:barChartDigitalLibraryDistribution}, we can see that among the reputed publishers, IEEE has twice as much work in this space as the others combined. 
However, the `other' and `arXiv' categories combined are higher, indicating considerable early-stage and unpublished works, also indicating the relative age and immaturity of this area of study. %

\subsubsection{Publications That Satisfy Selection Criteria}
Figure~\ref{fig:stackedBarIncExc} shows the distribution of papers that satisfy the selection criteria described in Table~\ref{tab:inclusion_and_exclusion_criteria_tab}. The GenAI-based works in the various PCB design and test life cycle stages broadly fall under LLMs, VLMs, GANs and diffusion models. The number of papers included is 80, which includes references to electrical circuit improvements and other subdivisions of electronics that apply to PCBs as well.

\begin{figure}[htbp]
  \centering
  \begin{minipage}{0.48\textwidth}
    \centering
    \resizebox{0.95\linewidth}{!}{\begin{tikzpicture}
    \begin{axis}[
        xbar stacked, %
        bar width=8mm, %
        width=12cm, %
        height=4cm, %
        xmin=0, xmax=227, %
        xlabel={Number of papers}, %
        symbolic y coords=\rotatebox{90}{Selected}, %
        ytick=data, %
        nodes near coords,
        nodes near coords align={center},
        every node near coord/.append style={
          color=black,        %
          inner sep=1pt
        },
        enlarge y limits=0.5,
        legend style={at={(0.35,1.275)}, anchor=north, legend columns=-1, draw=none}, %
        legend image post style={draw=none, mark=none},
    ]
        \addplot [forget plot, fill=green!25] coordinates {(80,\rotatebox{90}{Selected})};
        \addplot [forget plot, fill=red!25] coordinates {(147,\rotatebox{90}{Selected})};
        \addlegendimage{empty legend}
        \addlegendentry{Included / }
        \addlegendimage{empty legend}
        \addlegendentry{Excluded}
    \end{axis}
\end{tikzpicture}}
    \caption{Breakdown of the literature screening results according to inclusion and exclusion criteria.}
    \label{fig:stackedBarIncExc}
  \end{minipage}\hfill
  \begin{minipage}{0.48\textwidth}
    \centering
    \resizebox{0.95\linewidth}{!}{\begin{tikzpicture}
    \begin{axis}[
        xbar stacked, %
        bar width=8mm, %
        width=12cm, %
        height=4cm, %
        xmin=0, xmax=37, %
        xlabel={Number of papers}, %
        symbolic y coords=\rotatebox{90}{Stage-wise}, %
        ytick=data, %
        nodes near coords,
        nodes near coords align={center},
        every node near coord/.append style={
          color=black,        %
          inner sep=1pt
        },
        enlarge y limits=0.5,
        legend style={at={(0.50,1.275)}, anchor=north, legend columns=-1, draw=none}, %
    ]
        \addplot [fill=cyan!25] coordinates {(5,\rotatebox{90}{Stage-wise})};
        \addplot [fill=purple!25] coordinates {(16,\rotatebox{90}{Stage-wise})};
        \addplot [fill=gray!25] coordinates {(5,\rotatebox{90}{Stage-wise})};
        \addplot [fill=orange!25] coordinates {(1,\rotatebox{90}{Stage-wise})};
        \addplot [fill=green!25] coordinates {(8,\rotatebox{90}{Stage-wise})};
        \addplot [fill=red!35] coordinates {(2,\rotatebox{90}{Stage-wise})};
        \legend{Design spec., Circuit Sch. or Net., Layout, Printing, Assembly, Installation}
    \end{axis}
\end{tikzpicture}}
    \caption{Stage-wise distribution of the selected papers within the PCB design and test life cycle outlined in Figure~\ref{fig:pcb-supply-chain-life-cycle}.}
    \label{fig:stackedBarDesignStages}
  \end{minipage}
\end{figure}

\subsubsection{Distribution of Selected Papers in PCB Life Cycle} 

Although a total of 80 papers were retained after applying the selection criteria from Table~\ref{tab:inclusion_and_exclusion_criteria_tab}, only 37 of these could be precisely mapped to a specific stage within the PCB design and test life cycle. These are categorised in Figure~\ref{fig:stackedBarDesignStages}, which broadly classifies them according to the PCB design and test stage they fall under.

The remaining publications were still included in the survey as GenAI methodologies, auxiliary optimisation strategies, or adjacent problem spaces that influence the PCB life cycle indirectly are of significance to this survey. 
For example: analog circuit design is one component of PCB design that is essential for powering up and operating the board successfully. 

Works that focused on analog circuit design or mixed circuit design were also incorporated within the survey.

\section{Taxonomy and Application}
\label{sec:applying-taxonomy}

The taxonomy we propose in this paper is a classification method for each stage of the PCB design and test life cycle based on the objective of the proposed work. Since GenAI based work is aimed at automation, the three categories in which each PCB life cycle stage can be modified are: \textbf{(a) Design:} whether the work improves or automates part of the design generation step, \textbf{(b) Optimisation:} whether the selected work tweaks the performance (not time optimisation/reduction in turn-over times) of the PCB life cycle stage, and \textbf{(c) Validation:} whether the proposed work cosmetically and functionally verifies the PCB stage -- Functional Validation (Vf) -- or is able to conduct security checks, Security Validation (Vs), at that level.  

The identified works are classified against the taxonomy presented in Table~\ref{tab:taxonomy_table}. For every work, the stage of the PCB design and test life cycle which acts as an \textit{input} to the GenAI model is represented as \rotatebox[origin=c]{180}{\ding{119}}. The \textit{output} stage is represented as \ding{119}. For works that involve optimisation, iterative design generation or validation, the PCB stages can be \textit{both inputs and outputs} to the model. These stages are represented by \ding{108}. 

From the work we have assessed, LLMs sum up a majority of contributions while other GenAI-based work, like VLMs and GANs, are rampant in certain stages and under-researched in others.

\begin{table*}[tp]
    \centering
    \caption{The proposed taxonomy according to which all of the collected GenAI works, that directly relate to the various PCB manufacturing stages, can be classified. For each of the four categories: D, O, Vf and Vs, \rotatebox[origin=c]{180}{\ding{119}} represents the stage that is an input to the GenAI model, \ding{119} showcases the output stage and \ding{108} represents a stage that is both the input and output to the model.}
    \definecolor{Concrete}{rgb}{0.949,0.949,0.949}
    \resizebox{0.90\textwidth}{!}{\begin{tblr}{
  row{even} = {Concrete},
  cell{2}{1} = {r=6}{},
  cell{2}{3} = {r=6}{},
  cell{2}{4} = {r=6}{},
  cell{2}{5} = {r=6}{},
  cell{2}{6} = {r=6}{},
  vline{1-3,7} = {1-8}{},
  vline{3} = {3-8}{},
  hline{1-2,8} = {-}{},
}
Source & {PCB Lifecycle \\ Stage (can be multiple)} & Design (D) & Optimisation (O) &
{Functional\\Validation (Vf)} & {Security\\Validation (Vs)}\\
{The research \\paper in \\question. } & Specification (Spec.) &
{Does the work aim\\to automate design \\generation at \\this PCB stage?} &
{Does the work aim to \\assist with optimisation of\\ this PCB stage in terms\\ of power usage, performance, \\area, cost, or reduce \\resource usage on the PCB?} &
{Does the work target\\functional verification of this\\ PCB stage against its \\design specification?} &
{Does the work target\\ security validation of this\\PCB stage?}\\
 & {Schematic (Sch.) /\\Netlist (Net.)} &  &  &  & \\
 & Layout (Lay.) &  &  & & \\
 & Printing (Prt.) &  &  &  & \\
 & Assembly (Ass.) &  &  & & \\
 & Installation (Inst.) &  &  &  & \\
\end{tblr}
}
    \label{tab:taxonomy_table}
\end{table*}

The definitions for each entry in the taxonomised table are detailed as follows:

\noindent \textbf{Source} The research paper in question.\\
\textbf{Spec.} GenAI-based application at the design specification stage. Since this is an early design stage, it mostly involves interpreting or analysing a text-based requirements document.\\
\noindent \textbf{Sch./Net.} GenAI-based application at the circuit schematic or netlist (eg. BOM) stage. This stage involves laying out the circuit diagram and potentially simulating the circuit with tools such as LTSpice to ensure connectivity. \\
\textbf{Lay.} GenAI-based application at the circuit layout or routing stage. This usually involves connection of wires and circuit simulations to ensure that the system will be functional.\\
\textbf{Prt.} GenAI-based application at the board printing stage. This is generally already automated by manufacturing equipment or PCB printers, etc. in this space. Certain tests also exist here, e.g. flying probe. \\
\textbf{Ass.} GenAI-based application at the assembly stage. This can involve component installation and other finer operations like soldering which complete the circuit board, but may also involve sensing of assembled parts (e.g. camera checks).\\
\textbf{Inst.} GenAI-based application at the firmware installation stage. Firmware installation and functional and unit testing are a part of this stage.

\subsection{Design Specifications}
\label{sec:applying-taxonomy:designSpec}
LLMs have been demonstrated as effective in interpreting text-based inputs or natural language instructions. Therefore, for the design specification stage, which usually involves following a document that details design requirements and guidelines, LLMs have a significant and emerging capability compared to older automation methods \cite{zhao_special_2025}. 

In this early stage of PCB design and testing, much of the same principles as that of other electronics systems apply. Table \ref{tab:taxonomised_DesignSpec} highlights the various sources addressing LLM-based automation in the design specification stage and illustrates whether LLMs aid in automated design, optimisation or validation. 

\begin{table}[h]
    \centering
    \definecolor{Concrete}{rgb}{0.949,0.949,0.949}
    \caption{Design specification stage of the PCB design and test life cycle.}
    \resizebox{0.48\textwidth}{!}{
    \begin{tblr}{
  row{even} = {Concrete},
  cells = {c},
  vline{2,8} = {-}{},
  hline{1,6} = {-}{0.08em},
  hline{2} = {-}{},
}
Source & Spec. & Sch./Net. & Lay. & Prt. & Ass. & Inst. & D & O & Vf & Vs\\
{\cite{xiao_prefixllm_2024}} & \rotatebox[origin=c]{180}{\ding{119}} & \ding{108} &  &  &  &  & \ding{51} & \ding{51} & &\\
{\cite{shi_amsnet-kg_2025}} & \rotatebox[origin=c]{180}{\ding{119}} & \ding{108} &  &  &  &  & \ding{51} & \ding{51} & &\\
{\cite{lai_analogcoder_2025}} & \rotatebox[origin=c]{180}{\ding{119}} & \ding{119} &  &  &  &  & \ding{51} &  & &\\
\cite{liu_physics-informed_2024} & \ding{108} & \ding{119} &  &  &  &  &  &  & \ding{51} &\\
\end{tblr}

    \label{tab:taxonomised_DesignSpec}}
\end{table}

\subsubsection{Design: }

\cite{xiao_prefixllm_2024},\cite{shi_amsnet-kg_2025} and \cite{lai_analogcoder_2025} all convert design specifications documents into circuit netlists using LLMs. \cite{xiao_prefixllm_2024} showcases how LLMs can be used in the design stages of prefix circuit synthesis. The method is error-prone due to the structural and computational constraints of digital logic circuits, as noted in \cite{xiao_prefixllm_2024}. In contrast, \cite{shi_amsnet-kg_2025} and \cite{lai_analogcoder_2025} are implementations for analog circuit design. \cite{shi_amsnet-kg_2025} showcases a knowledge graph-based method to translate connections between various aspects of the circuit in its netlist representation, along with two block level netlists in its dataset, while \cite{lai_analogcoder_2025} converts design specifications into Python code using LLMs, and modifies the resultant code to netlists. Despite \cite{shi_amsnet-kg_2025} and \cite{lai_analogcoder_2025} being analog circuit improvements, the methods in which LLMs are applied to their design stages are worth noting, and could easily play a role in circuit design and netlist design for PCBs, which is covered in Section~\ref{sec:applying-taxonomy:circuitSchAndNet}.

\subsubsection{Optimisation: }
\textit{Retrieval-augmented generation (RAG)} is a GenAI technique that provides context to its models to enhance quality of outputs.
\cite{xiao_prefixllm_2024} applies an iterative refinement strategy with \textit{chain of thought} prompting, a learning approach for LLMs to improve their performance in reasoning based tasks \cite{wei_chain--thought_2023}.
A contrasting line of work, \cite{shi_amsnet-kg_2025}, adopts Bayesian Optimisation to determine optimal parameter configurations for a KG-RAG topology, and a shift towards algorithmic and parameter-level optimisation. These differing works highlight a divergence in current LLM optimisation research for circuit design, where efforts are split between enhancing LLM reasoning mechanisms and improving computational efficiency of circuit design itself through parameter tuning.

\subsubsection{Validation: }
\cite{liu_physics-informed_2024} proposes LP-COMDA, an LLM-power machine learning planner that validates designs against its specifications through an interactive chat-based interface. The framework comprises of two neural network models that simulate and learn switch circuit behaviours. While the ML part of the model performs design generation and optimisation, the LLM aspect or rather GenAI aspect only deals with validation of the created circuit against its provided specifications.

From our taxonomy for the design specification stage described in Table~\ref{tab:taxonomised_DesignSpec}, we note that manual intervention or validation by a human resource is still a necessity, as only one work touches on LLM-based validation of design specifications against generated designs. Even in \cite{liu_physics-informed_2024}, the framework is a chatbot, which implies iterative prompts and feedback are collected by the person using it.

\begin{takeaway}
For the design specification stage of PCB design, our survey finds that security vailidation is an area where GenAI has yet to be applied. Further, only one work has explored functional validation -- which was interactive, requiring a human engineer to provide inputs and feedback to their GenAI framework. There exists a gap here for security and autonomous functional validation.
\end{takeaway}

\subsection{Circuit Schematic/Netlist}
\label{sec:applying-taxonomy:circuitSchAndNet}

Circuit schematics are visual representations of how the analog and digital components on the board are connected with each other. These diagrams are best interpreted and analysed using computer vision models and machine learning algorithms, but GenAI has demonstrated great success with LLMs and VLMs in this area, especially for analog circuits. 
In this subsection, we discuss the recent advancements in circuit schematic design or netlist generation in PCBs, along with a bird's eye focus on analog circuits, whose principles or methodologies could also apply to PCB contexts. 

\begin{table}[h]
    \centering
    \definecolor{Concrete}{rgb}{0.949,0.949,0.949}
    \caption{Circuit schematic/netlist stage of the PCB design and test life cycle.}
    \resizebox{0.48\textwidth}{!}{
    \begin{tblr}{
  row{even} = {Concrete},
  cells = {c},
  vline{2,9} = {-}{},
  hline{1-2,18} = {-}{0.08em},
}
Source & Spec. & Sch. & Net. & Lay. & Prt. & Ass. & Inst. & D & O & Vf & Vs\\
\cite{chang_lamagic_2024} & \rotatebox[origin=c]{180}{\ding{119}} &  & \ding{108} &  &  &  &  & \ding{51} & \ding{51} &  &\\
\cite{liu_ampagent_2024} &  & \ding{119} & \ding{108} &  &  &  &  & \ding{51} & \ding{51} & &\\
\cite{vijayaraghavan_circuitsynth_2024} &  & \ding{108} &  &  &  &  &  & \ding{51} & \ding{51} & &\\
\cite{lin_pe-gpt_2025} & \rotatebox[origin=c]{180}{\ding{119}} & \ding{108} & \ding{119} &  &  &  &  & \ding{51} & \ding{51} & & \\
\cite{shen_atelier_2025} & \rotatebox[origin=c]{180}{\ding{119}} & \ding{108} &  &  &  &  &  & \ding{51} & \ding{51} & \ding{51} & \\
\cite{raval_circuit_2025} &  &  & \ding{108} &  &  &  &  & \ding{51} &  & & \\
\cite{matsuo_schemato_2024} &  & \ding{108} & \rotatebox[origin=c]{180}{\ding{119}} &  &  &  &  & \ding{51} & \ding{51} & &\\
\cite{bhandari_masala-chai_2025} &  & \ding{119} & \rotatebox[origin=c]{180}{\ding{119}} &  &  &  &  & \ding{51} &  &  & \\
\cite{chaudhuri_spiced_2024} & \rotatebox[origin=c]{180}{\ding{119}} &  & \ding{108} & &  &  &  & \ding{51} &  & \ding{51} & \\
\cite{liu_ladac_2024} & \rotatebox[origin=c]{180}{\ding{119}} & \ding{108} &  &  &  &  &  & \ding{51} & \ding{51} & \ding{51} &\\
\cite{s_llm-uso_2025} &  & \ding{108} & \ding{119} &  &  &  &  &  & \ding{51} &  &\\
\cite{shaik_using_2025} & \rotatebox[origin=c]{180}{\ding{119}} & \ding{108} &  &  &  &  &  & \ding{51} &  & & \\
\cite{shi_amsbench_2025} & \rotatebox[origin=c]{180}{\ding{119}} & \rotatebox[origin=c]{180}{\ding{119}} & \rotatebox[origin=c]{180}{\ding{119}} &  &  &  &  &  &  & \ding{51} & \\
\cite{nau_spiceassistant_2025} & \rotatebox[origin=c]{180}{\ding{119}} & \ding{108} & \rotatebox[origin=c]{180}{\ding{119}} &  &  &  &  & \ding{51} &  & \ding{51} & \\
\cite{liu_llm-based_2025} & \rotatebox[origin=c]{180}{\ding{119}} & \ding{108} & \ding{119} &  &  &  &  &  & \ding{51} & \ding{51} &\\
\cite{ecik_anomaly_2023} & \rotatebox[origin=c]{180}{\ding{119}} & \rotatebox[origin=c]{180}{\ding{119}} & \ding{119} &  &  &  &  &  &  & \ding{51} & \\
\end{tblr}

    \label{tab:taxonomised_SchemNetlist}}
\end{table}

Table \ref{tab:taxonomised_SchemNetlist} lists all these sources along with whether they are used in design stages, optimisation stages or validation stages. 

\subsubsection{Design:}
\textbf{Reasoning LLMs:}
LLMs are particularly useful in iteratively analysing and improving their outputted designs through chain-of-thought reasoning. 
In the analog space, \cite{liu_ampagent_2024} applies multi-agent reasoning LLMs to text-based circuit netlists and designs complex amplifier schematics for analog circuits iteratively. 
\cite{liu_ladac_2024} takes in a specification in the form of a prompt, and generates an analog circuit design as well as a simulation for the same. 
These two works show design generation for analog components and circuits using interactive LLMs and chain-of-thought reasoning. 
But LLMs are not limited to these applications. 
They also effective in \textbf{classification-based tasks}, such as \cite{vijayaraghavan_circuitsynth_2024}, which creates a circuit dataset and uses a validity classifier to generate its circuit topologies. 
\cite{chang_lamagic_2024} also creates graph-based representations of circuits with a labelled dataset, a technique called supervised learning, to generate circuit designs, illustrating that classification methods are equally useful avenues towards circuit design automation. 
While none of these works directly focus on PCBs, their application usages are helpful to note and can be applied to PCB contexts as well, leveraging both the reasoning and classification expertise of LLMs to automate parts of the PCB life cycle.

\textbf{Knowledge based: }
It is essential to understand that good-quality LLM generated outputs also require clear and concise inputs being fed into them. Representation methods like symbolic knowledge graphs, as noted in \cite{shaik_using_2025}, have proven to be useful on LLMs to establish relationships between the various analog components in design, and generate circuit schematics as a result. \cite{shen_atelier_2025} is another method that curates a knowledge base using retrieval-augmented generation (RAG) and generates analog circuit designs with the help of multiple LLM agents, where each agent specialises in a different stage of circuit design. 
This multi-agent framework has an advantage over general purpose LLMs and other black-box circuit generation methods. However, PE-GPT \cite{lin_pe-gpt_2025} further extends LLM capabilities by proposing a multimodal LLM tailored for power electronics design, which combines RAG and metaheuristic algorithms. 
The prompt is fed with design instructions and domain-centric knowledge which enhances its reasoning capabilities and improves the quality of generated outputs, in comparison to other state-of-the-art LLMs. 

\textbf{SPICE Netlists: }
At the netlist level, \cite{raval_circuit_2025} is a framework which fine-tunes GPT-4 and proposes Circuit AI, that automatically retrieves the required part from the design's bill of materials.
For components past their end-of-life, or where a substitution part is required, such an automated approach is beneficial. 
Meanwhile, for analog circuits, \cite{matsuo_schemato_2024} fine-tunes an LLM, Llama 3.18B, with various analog components that are used in circuit design and, as a result, the framework is capable of generating LTSpice schematics from SPICE netlists. 
A similar framework that converts SPICE netlists to circuit topologies is \cite{bhandari_masala-chai_2025}, but uses some machine-learning algorithms to assist LLMs in design generation.
\cite{nau_spiceassistant_2025} generates adapted netlists that fit the inputted user specifications as well as interactive feedback from the user at each LLM netlist output stage. 
SPICED \cite{chaudhuri_spiced_2024} is an LLM-based tool that helps generates netlists for analog circuits. 
All these works incorporate LLMs but while \cite{raval_circuit_2025} is a PCB focused improvement, the other four works focus on analog or mixed circuit design.

\subsubsection{Optimisation:}
\cite{matsuo_schemato_2024} finetunes Llama 3.18B with LTSpice netlist-schematic key-value pairs, but the only drawback is that it lacks structural diversity due to its limited training data. Meanwhile, in \cite{lin_pe-gpt_2025} and \cite{shen_atelier_2025}, the frameworks optimise design parameters to generate the most optimal design for the provided specifications where the former uses a multimodal LLM and the latter multiple LLM agents. 

A particular step within \cite{vijayaraghavan_circuitsynth_2024} focuses on refining circuit topologies using LLMs.
Similarly, an LLM-based AI agent is used in \cite{liu_llm-based_2025} to integrate with circuit simulators and data analysis tools and automate AMS transistor sizing. Both these works optimise only one aspect (i.e. just the refinement or just the transistor sizing) of the circuit. This showcases how identifying which part of the design or test life cycle benefits the most from automation changes the methodology drastically.   

\cite{liu_ladac_2024} and \cite{s_llm-uso_2025} optimise circuits using a swarm algorithm called Artificial Bee Colony (ABC) and a probabilistic algorithm called Bayesian Optimisation respectively to create text-based representations of circuits with LLMs. 
This difference in algorithm usage is attributed to the varying problem areas in the models. 
The former provides a scalability solution that can benefit larger analog circuits as it automatically optimises the generated circuit without expecting human instructions, while the latter improves computational performance of the LLM rather than the generated circuit.

\subsubsection{Validation:}
\cite{shi_amsbench_2025} compares and contrasts multi-modal LLMs and how they perform against 8000 different analog or mixed circuit problems. 

In terms of SPICE simulation generation, \cite{shen_atelier_2025}, \cite{liu_ladac_2024} and \cite{liu_llm-based_2025} self-validate their generated designs to ensure that the outputted design is a closed circuit as expected. Similarly, \cite{nau_spiceassistant_2025} proposes SPICEAssistant, an LLM-based SPICE tool that improves switched-mode power supply (SMPS) circuit design by enabling iterative SPICE simulations. On the other hand, anomaly detection or malicious behaviour can also be detected using simulations. \cite{ecik_anomaly_2023} showcases how signal integrity checks for EMCs can be done using simulations, and anomalies can be detected with the help of internal decision trees. All the above mentioned validation techniques require intervention from humans to ensure circuit behaviour is as intended. 

One method to detect analog Trojans and syntactical bugs with minimal human effort in circuit netlists is with SPICED \cite{chaudhuri_spiced_2024}, which is an LLM-powered framework. It can be considered a defensive strategy for circuits, by identifying malicious modifications early on in the PCB design and test life cycle. 

\begin{takeaway}
The circuit schematic and netlist generation stages also have not applied GenAI models to verify or detect security vulnerabilities in designs. While a significant number of functional validation-based GenAI works are proposed at this stage, most of them involve GenAI creating SPICE simulations which require a human to interactively validate them. While more works here have GenAI producing and optimising designs, no autonomous functional or security relevant validation frameworks are presented.
\end{takeaway}

\subsection{Layout and Routing}

The layout and routing stage of the PCB design and test life cycle includes determining the optimal space and orientation of the various components, as well as the most efficient way they can be routed. The surface constraints, spatial irregularities and boundary restrictions in PCBs affect how components can be oriented or wired on the board  \cite{chen_pcbagent_2025}. In this section, we explore how GenAI is used in this space and how it overcomes the current challenges of PCB layout and routing.

Table \ref{tab:taxonomised_LayoutRouting} lists recent works in this area and whether GenAI is used in design, optimisation or validation stages.

\begin{table}[h]
    \centering
    \definecolor{Concrete}{rgb}{0.949,0.949,0.949}
    \caption{Layout and routing stage of the PCB design and test life cycle. 
    }
    \resizebox{0.48\textwidth}{!}{
    \definecolor{Concrete}{rgb}{0.949,0.949,0.949}
\begin{tblr}{
  row{even} = {Concrete},
  cells= {c},
  vline{2,8} = {-}{},
  hline{1-2, 7} = {-}{0.08em},
}
Source & Spec. & Sch./Net. & Lay. & Prt. & Ass. & Inst. & D & O & Vf & Vs\\
\cite{you_interactive_2024} &  & \rotatebox[origin=c]{180}{\ding{119}} & \ding{108} &  &  &  & \ding{51} & \ding{51} & \ding{51} &\\
\cite{chen_pcbagent_2025} & \rotatebox[origin=c]{180}{\ding{119}} &  & \ding{108} &  &  &  & \ding{51} & \ding{51} & & \\
\cite{zhang_applying_2025} & \rotatebox[origin=c]{180}{\ding{119}} &  & \ding{119} &  &  &  & \ding{51} &  & \ding{51} &\\
\cite{chen_llm-enhanced_2024} & & \rotatebox[origin=c]{180}{\ding{119}} & \ding{119} &  &  &  &  & \ding{51} & & \\
\cite{li_functional_2024} &  &  & \rotatebox[origin=c]{180}{\ding{119}} &  &  &  &  &  & \ding{51} & \\
\end{tblr}

    \label{tab:taxonomised_LayoutRouting}}
\end{table}

\subsubsection{Design:}
\cite{you_interactive_2024} is a framework powered by an LLM (ChatGPT-4) that looks at a SPICE netlist, and generates placement and routing commands in code. This framework leverages LLM and its expertise with generating code for HDL \cite{thakur_autochip_2024}, and other software problems. 
In contrast, \cite{zhang_applying_2025} and \cite{chen_pcbagent_2025}  do not output any code. 
\cite{zhang_applying_2025} proposes a few-shot and chain-of-thought LLM-based approach to generate necessary routing coordinates and assist with complex PCB routing tasks. 
While \cite{zhang_applying_2025} is an LLM only framework, \cite{chen_pcbagent_2025} is an ML-based layout automation framework that uses LLMs interactively to improve an ML-generated placement result against its design specifications. 
It is therefore is more powerful than an LLM only approach, and proves that it outperforms state-of-the-art methods in 17 real-world tasks.

\subsubsection{Optimisation:}
\cite{you_interactive_2024} uses ChatGPT-4 interactively on SPICE netlists to satisfy the necessary design constraints, and ensure placement is strategically performed for various mixed circuit applications.
Meanwhile, \cite{chen_pcbagent_2025} also uses LLMs interactively to ensure industrial constraints are met when compared to design specifications. 
In the analog layout space, LLANA \cite{chen_llm-enhanced_2024} improves layout synthesis by applying the Bayesian Optimisation method to analog net weighting constraints. 
It is worth noting that all three frameworks directly optimise the layout stage or operations performed in the layout stage, and not the LLM performing the automation.

\subsubsection{Validation:}
Traditional methods of layout validation include layout versus schematic checks and sometimes design rule checks \cite{you_interactive_2024}.
However, in Table~\ref{tab:taxonomised_LayoutRouting}, LLMs are used in \cite{you_interactive_2024} with human intervention, to ensure that the generated circuit designs are correct. 
\cite{zhang_applying_2025} validates synthetic and fine-tuned examples of PCB layouts with a validation expert iteratively prompting the LLM. 
The above two methods apply LLMs to the design validation step of the layout stage while in \cite{li_functional_2024}, circuit layouts are validated with GANs, to observe PCB electrical pathways and detect anomalies. 
None of these GenAI based approaches are end-to-end automation frameworks. 

\begin{takeaway}
At the layout, placement and routing stage, security validation continues to be an overlooked area. Functional validation models use iterative LLMs and GANs models to validate the generated layout designs, but a validation expert is expected to confirm the final design files. Even at this stage, no GenAI tools independently validate PCB layouts -- they instead work as EDA design assistants.
\end{takeaway}

\subsection{Printing}
The PCB printing stage constitutes of the fabrication of the board substrate. 
These include applying the photo-resistive layer, trace etching, drilling, plating, solder mask application and silkscreen printing. The output of this stage is an etched and drilled board. 
Table~\ref{tab:taxonomised_Printing} classifies this work against our taxonomy.

\begin{table}[h]
    \centering
    \definecolor{Concrete}{rgb}{0.949,0.949,0.949}
    \caption{Printing stage of the PCB design and test life cycle.}
    \resizebox{0.48\textwidth}{!}{
    \begin{tblr}{
  row{even} = {Concrete},
  vline{2,8} = {-}{},
  cells = {c},
  hline{1,3} = {-}{0.08em},
  hline{2} = {-}{},
}
Source & Spec. & Sch./Net. & Lay. & Prt. & Ass. & Inst. & D & O & Vf & Vs\\
\cite{gang_character_2021} &  &  & \rotatebox[origin=c]{180}{\ding{119}} & \rotatebox[origin=c]{180}{\ding{119}} &  &  &  &  & \ding{51} & \\
\end{tblr}

    \label{tab:taxonomised_Printing}}
\end{table}

\subsubsection{Validation:}
\cite{gang_character_2021} specifically focuses on character recognition on PCBs and uses a convolutional neural network in conjunction with GANs to identify these imprinted characters accurately. 
Though this framework is a CNN-based approach applied at the assembly and final inspection stages, the printing stage is responsible for issues or anomalies in the engraved characters. Therefore, it is classified in this area. 

To our knowledge, this is the only GenAI work that impacts the PCB printing stage.

\begin{takeaway}
While machine learning based methods and computer vision frameworks have been applied in the domain of PCB printing, few GenAI-based works exist in this area. However, \cite{gang_character_2021} indicates that adopting GenAI for PCB verification at the printing stage should be possible. Here, too, no efforts have yet been made to validate security threats or risks that could arise while printing.  
\end{takeaway}

\subsection{Assembly}
Assembly for PCBs involves attaching components to the board, soldering, component placement and quality inspections. 
The result of this stage is a ready-to-use PCB that has not been configured to any application yet. 
At this stage, we consider board manufacturing to be complete and hence, none of the GenAI works output an assembled board. 

All the works identified in this PCB design and test stage are design-validation based methods that identify defects on the board. Design creation and optimisation are not a concern at this stage, as all of the manufacturing is completed. The identified design and optimisation classification gaps are not worth exploring for this reason. 

Table~\ref{tab:taxonomised_Assembly} classifies these PCB defect detection works.

\begin{table}[h]
    \centering
    \definecolor{Concrete}{rgb}{0.949,0.949,0.949}
    \caption{Assembly stage of the PCB design and test life cycle.}
    \resizebox{0.48\textwidth}{!}{
    \begin{tblr}{
  row{even} = {Concrete},
  cells = {c},
  vline{2,8} = {-}{},
  hline{1,10} = {-}{0.08em},
  hline{2} = {-}{},
}
Source & Spec. & Sch./Net. & Lay. & Prt. & Ass. & Inst. & D & O & Vf & Vs \\
\cite{ilchuk_m_novel_2023} & \ding{108} &  &  & \rotatebox[origin=c]{180}{\ding{119}} & \rotatebox[origin=c]{180}{\ding{119}} &  &  &  & \ding{51} & \\
\cite{liu_printed_2025} &  & \ding{119} &  & \rotatebox[origin=c]{180}{\ding{119}} & \rotatebox[origin=c]{180}{\ding{119}} &  &  &  & \ding{51} & \\
\cite{huang_defect_2025} &  & \ding{119} &  &  & \rotatebox[origin=c]{180}{\ding{119}} &  &  &  & \ding{51} & \\
\cite{wang_conditional_2023} &  & \ding{119} & \rotatebox[origin=c]{180}{\ding{119}} & \rotatebox[origin=c]{180}{\ding{119}} & \rotatebox[origin=c]{180}{\ding{119}} &  &  &  & \ding{51} & \\
\cite{you_pcb_2022} &  &  &   &  & \rotatebox[origin=c]{180}{\ding{119}} &  &  &  & \ding{51} & \\
\cite{luo_defectdiffuser_2025} &  &  & \rotatebox[origin=c]{180}{\ding{119}} & \rotatebox[origin=c]{180}{\ding{119}} & \rotatebox[origin=c]{180}{\ding{119}} &  &  &  & \ding{51} & \\
\cite{xu_printed_2025} &  &  &  &  & \rotatebox[origin=c]{180}{\ding{119}} &  &  &  & \ding{51} & \\
\cite{liu_automatic_2024} &  &  &  &  & \rotatebox[origin=c]{180}{\ding{119}} &  &  &  & \ding{51} & \\
\end{tblr}

    \label{tab:taxonomised_Assembly}}
\end{table}

\subsubsection{Validation: }
Visual LLMs are backed by machine learning algorithms. They fare well in PCB defect detection due to their image detection and recognition capabilities. In \cite{ilchuk_m_novel_2023}, it is used as an interactive tool to detect PCB defects with a computer vision based framework in the form of prompts to the LLM agent. 
As a result, this uses an image-to-text generator to map the design to natural language, which traditional LLMs do not support. 
This work validates the design specifications through inspecting the PCB as images, and hence falls under the umbrella of VLMs. 

GANs also predominates this space in PCB defect detection and integrates well with machine learning algorithms to handle image-based design validation. \cite{liu_printed_2025} is one such framework that studies the defect detection process in PCB assembly and welding, using a computer vision model enhanced by GANs to increase synthetic training data. 
Likewise, \cite{huang_defect_2025}, a CNN-based approach that uses GANs to generate synthetic anomaly samples, focuses on identifying six common defect types: missing hole, rat bite, open circuit, short circuit, burr, and virtual welding with the help of an enhanced YOLOv11 model. 
However, there are also works that only deal with augmenting data samples. \cite{wang_conditional_2023} creates a diversified anomaly dataset by inspecting the routing on PCBs and generating synthetic data with those samples. The dataset produced by \cite{wang_conditional_2023} is tested by the cTransGAN framework against two existing PCB datasets to determine how comprehensive it is in generating PCB anomaly samples. While \cite{liu_printed_2025} and \cite{huang_defect_2025} can be used directly to inspect PCBs, \cite{wang_conditional_2023} is an augmentation strategy that is tested within the work, validating that other inspection models can be pre-trained with it.  

Based on the categorisation of PCB defects in \cite{alawandi_pcb_2025}, ten defects fall under the umbrella of bare-PCB defects. Of these, six are identified by a framework called EESRGAN \cite{you_pcb_2022}, which is a GAN-based data pre-processing approach that uses PCB pictures from existing datasets (for eg: the HRIPCB dataset), and enhances them for PCB defect recognition. Similarly, \cite{luo_defectdiffuser_2025} proposes a framework called DEFectDiffuser, that uses a diffusion model to generate realistic PCB defect images. It does so by combining low-resolution defects with high-resolution textures to address sample scarcity. \cite{luo_defectdiffuser_2025} also helps identify design flaws against the same six groups of defects. The differences between the two are \cite{you_pcb_2022} is a GAN based approach applied to existing datasets, and \cite{luo_defectdiffuser_2025} integrates the augmentation as part of its pre-trained diffusion model. 

Between \cite{huang_defect_2025}, \cite{you_pcb_2022} and \cite{luo_defectdiffuser_2025}, the former notes virtual welding defects while the latter two identify spurious copper design defects, that distinguish them from each other.

Another approach to defect detection involves Stable Diffusion models, which deal with text inputs to image outputs and apply denoising on them. \cite{xu_printed_2025} is one such method that automatically creates PCB defect samples by combining ControlNet and a Stable Diffusion Model to detect soldering defects. \cite{liu_automatic_2024} uses the same approach to detect PCB features using localisation algorithms and segmentation. Both these works focus on flaws targeting the net level and component level respectively.

\begin{takeaway}
By the assembly stage, most of the board manufacturing is complete and any GenAI based work at this stage targets defect detection at some level. However, like previous stages, security validation remains unexplored.
\end{takeaway}

\subsection{Installation} \label{sec:applying-taxonomy:installation}
By this stage of the PCB design and test life cycle, manufacturing of the board is complete. 
Connecting peripheral devices, firmware installation and other application dependent work fall into this area as the output of this stage is a fully functional system that is ready to be distributed.
Embedded systems or cyber-physical systems can also be taxonomised at this stage as they are fully functional systems programmed to perform a particular function.

At the same time, this is the last stage in which security flaws can be detected and fixed before end users get their hands on the devices. 
Therefore, hardware security based work that does not fall under any of the PCB design and test life cycle stages is identified and taxonomised in this section. 

All previously identified works that make up a completely functional system are classified in Table~\ref{tab:taxonomised_Installation}.

\begin{table}[h]
    \centering
    \definecolor{Concrete}{rgb}{0.949,0.949,0.949}
    \caption{Installation stage of the PCB design and test life cycle. At this stage, the system is akin to an embedded system or cyber-physical system as it is a usable hardware device.}
    \resizebox{0.48\textwidth}{!}{
    \begin{tblr}{
  row{even} = {Concrete},
  cells = {c},
  vline{2,8} = {-}{},
  hline{1,9} = {-}{0.08em},
  hline{2} = {-}{},
}
Source & Spec. & Sch./Net. & Lay. & Prt. & Ass. & Inst. & D & O & Vf & Vs \\
\cite{englhardt_exploring_2024} & \rotatebox[origin=c]{180}{\ding{119}} &  &  &  &  & \ding{108} & \ding{51} & \ding{51} & \ding{51} & \\
\cite{sabree_openai_2024} & \rotatebox[origin=c]{180}{\ding{119}} &  &  &  &  & \ding{119} & \ding{51} &  & & \\
\cite{jansen_words_2023} & \rotatebox[origin=c]{180}{\ding{119}} & \ding{119} &  &  &  & \ding{119} & \ding{51} &  & \ding{51} & \\
\cite{lu_enabling_2024} & \rotatebox[origin=c]{180}{\ding{119}} & \ding{108} & \ding{108} &  &  & \ding{119} & \ding{51} & \ding{51} & \ding{51} & \\
\cite{cui_large_2024} & \rotatebox[origin=c]{180}{\ding{119}} &  &  &  &  & \ding{119} & \ding{51} & \ding{51} & \ding{51} &  \\
\cite{kokolakis_harnessing_2024} & \rotatebox[origin=c]{180}{\ding{119}} & \rotatebox[origin=c]{180}{\ding{119}} &  &  &  & \ding{108} &  &  &  & \ding{51} \\
\cite{gohil_llmpirate_2024} &  & \ding{108} &  &  &  & \rotatebox[origin=c]{180}{\ding{119}} & \ding{51} &  &  & \ding{51} \\
\end{tblr}

    \label{tab:taxonomised_Installation}}
\end{table}

\subsubsection{Design: }
In the EDA space, \cite{lu_enabling_2024} presents an interactive and iterative LLM-based tool for embedded mechanical computation design that generates the circuit netlist as a text-based intermediate stage, proposes a circuit schematic and describes and optimises the best possible layout for design. 
Since \cite{lu_enabling_2024} focuses on the ideation of a fluidic computation interface, it does not deeply dive into the printing and assembly stages of system manufacturing, and instead elaborates more into the implementation of the proposed interface.
While it is not a completely autonomous approach, such a tool reduces the manual effort involved in proposing high-level designs according to the inputted design specification.
This in turn makes for faster design iteration cycles, an improvement to the existing embedded system manufacturing life cycle. 

Another example of using LLMs to generate high level designs is in \cite{jansen_words_2023}, that prompts the model with some easy electronics design problems and obtains netlist representations of circuit specifics as well as LLM-generated micro-controller code. This work is a promising avenue for EDA, but its experiments are performed on an already assembled electronic system that has been created and validated by a human. Therefore, the work does not automate any of the PCB design and test life cycle stages.  

Interactive LLMs are also used in embedded systems programming to generate code, like in \cite{sabree_openai_2024} which likely uses at least ChatGPT-4 considering the date of publishing. The exact model is not described in the work. \cite{englhardt_exploring_2024} takes code generation using LLMs a step further by also debugging and finding syntactical errors in embedded system programming. 

In terms of black box design manipulations, \cite{gohil_llmpirate_2024} proposes an end-to-end LLM framework that takes in netlist representations of a circuit and maliciously modifies them in a way that detection tools, that have no access to design files (hence black box) but are used to validate manufactured designs cannot determine design flaws.

Finally, \cite{cui_large_2024} applies a multi-agent LLM-based framework to automate controller design in power electronics. The advantage of a multi-agent framework is different tedious design tasks can be channeled to different agents such as one for modelling, one for generating the algorithm and one for optimisation. 

\subsubsection{Optimisation: }
According to our taxonomy, GenAI models can be either optimised specifically to automate one of the PCB design and life cycle stages -- which includes enhancing performance of the stage itself -- or to tweak parameters of the GenAI model to improve overall automation performance at the applied PCB stage. 

For example, optimisation in \cite{lu_enabling_2024} involves internally iterating over its proposed netlist, schematic and layout instructions based on asynchronous user prompts.
Similarly, \cite{englhardt_exploring_2024} optimises its own work as its goal is to detect syntactical bugs in the code generated by the LLM. Both of these works automate the previously generated design using GenAI models, specifically LLMs.

However, \cite{cui_large_2024} performs optimisation with the goal of tweaking the overall system parameters through an interactive agent, that takes inputs from the user on which of the 2 optimisation algorithms to use. Considering just the context of this LLM agent in the multi-agent framework, the inputs to it are prompt instructions determined by the user on which optimisation algorithm to pick. Internally, it will use either the Particle Swarm Optimisation algorithm or an adaptive optimisation strategy called Genetic Algorithm based on the inputted choice. The end result of \cite{cui_large_2024} is an automatically optimised controller design. 

\subsubsection{Validation: }
A validation engineer can interact with \cite{englhardt_exploring_2024} to identify issues in the ChatGPT-4 generated code. Similarly, \cite{jansen_words_2023} uses two state-of-the-art LLMs to generate high level electronic designs and benchmarks them against its datasets: MICRO25 and PINS100. While the former applies an interactive validation approach with a human in the loop, \cite{jansen_words_2023} classifies whether or not simple tasks can be completed using binary classification methods again natural language instructions.

\textbf{Security: }
In terms of design implementation validation, or security validation in other words, \cite{kokolakis_harnessing_2024} showcases how LLMs can assist attackers in inserting hardware Trojans into complex CPU designs by identifying and modifying implementation modules. Similarly, LLMPirate \cite{gohil_llmpirate_2024} is the first LLM-based method to automatically generate pirated hardware circuit designs which evade state-of-the-art detection through IP piracy tools. 
In both \cite{kokolakis_harnessing_2024} and \cite{gohil_llmpirate_2024}, implementation flaws are introduced to PCB design through LLMs. While \cite{kokolakis_harnessing_2024} highlights an attacker strategy, \cite{gohil_llmpirate_2024} performs experiments that point out loopholes in existing board-level detection methods, which can be used to improve board-level defenses.

\begin{takeaway}
The installation stage is the first stage in the life cycle that notes GenAI based automation work in the security validation area, and two identified works highlight implementations in design that either maliciously intercept or evade existing anomaly detection strategies. However, no identified works combine functional validation and security validation simultaneously.
\end{takeaway}

\section{Discussions} \label{sec:discussions}
In this section we draw connections between the trends we've identified in domains adjacent to LLMs for PCBs with our proposed taxonomy, then elaborate on our key takeaways from this survey.

\subsection{Neighboring Domain Trends} \label{sec:discussions:neighboringDomainTrends}

\subsubsection{EDA and LLMs}
LLMs applied to EDA is closely linked to GenAI-based PCB automation, if not directly a superset of it, though as noted most works in this space focus on IC design rather than the other kinds of electronic hardware.
Exploring and identifying how GenAI is used in PCB design can serve to highlight what is currently lacking in the IC design space, as well as what the scope for future directions is in this domain.

\textbf{Design: }
From Table~\ref{tab:llm_eda_survey}, we see that LLMs in EDA are predominantly used for HDL code generation for integrated circuits. 
\textit{We identified 5 out of 6 collected LLM for EDA surveys focusing on `hardware design generation' steps (i.e. creating HDL code), but that do not investigate other aspects of integrated chip design, or other hardware system design operations as the phrase suggests.}

\textbf{Optimisation: }
Only 2 of the 6 surveys consider optimising existing EDA process through LLMs. 
Of these, one optimises integrated circuit generation and the other suggests methods from literature that can tweak existing EDA processes without implementing those changes on LLMs. 

\textbf{Validation: }
5 out of the 6 identified surveys have explored validation efforts in the context of LLM code generation, like bug detection using LLMs and the syntactical inspection of LLM-generated code. 

\textbf{Security: }
While Table~\ref{tab:llm_eda_survey} shows that 2 of the 6 surveys involve security, deeper inspection reveals that \cite{kande_llms_2024} and \cite{xu_llm-aided_2024} only explore risks and vulnerabilities that LLMs can pose in hardware design and HDL generation respectively. 
Neither surveys identify or propose unique methodology-based solutions to identify or solve a pre-existing security vulnerability on PCBs using GenAI. 

\begin{takeaway}
LLMs in EDA primarily focus on HDL code generation, but not other aspects of hardware system design. We also note that in the security validation domain, the surveys we evaluated explore and detail security risks that LLMs can pose in hardware design generation, but few novel methods are proposed which contribute to the security space.
\end{takeaway}

\subsubsection{Embedded and cyber-physical systems}
In Section~\ref{sec:applying-taxonomy:installation} (Installation), we infer that manufactured PCBs, with all their components attached and their firmware and peripherals connected, resemble embedded systems / cyber-physical systems. 
Analysing trends in these domains can help cross-validate expected versus actual observations within PCBs as well. Table~\ref{tab:relatedWorkcomparison} outlines 7 identified works within embedded systems and cyber-physical systems works. 
 
\textbf{Design:}
All 7 works are LLM-based design generation strategies. None of these recent collected works shed light on any other GenAI models. 
3 of these generate code or debug previously generated code with LLMs, \cite{lu_enabling_2024} and \cite{cui_large_2024} generate high level embedded system designs and controller designs, and the remaining 2 survey LLM applications in cyber-physical systems and IoT respectively. 

\textbf{Optimisation:}
Of the 7 collected works, 2 optimised parts of the design. 
While \cite{cui_large_2024} applies one of two optimisation algorithms to the problem depending on the inputted prompt, \cite{sarhaddi_llms_2025} discusses how design optimisation is performed for various IoT systems in the 260 articles it surveys.

\textbf{Validation:}
Amongst the 7 sources, in the functional validation space, only 3 attempt to validate generated designs using GenAI. 
\cite{englhardt_exploring_2024} debugs embedded systems code, \cite{jansen_words_2023} benchmarks LLM-generated code against 2 of its proposed benchmarks, and \cite{cui_large_2024} illustrates the use of multi-agent LLMs to generate designs. 
All three use LLMs to perform automation; again, none of these collected works touch on other GenAI models. 

\textbf{Security:} 
The two identified security validation based works, \cite{kokolakis_harnessing_2024} and \cite{gohil_llmpirate_2024}, exploit existing security issues and outline attacker strategies to evade current board-level defense methods. 
However, \textit{neither of these works propose methodology based solutions to counter the existing security vulnerabilities.} For this reason, these frameworks are not novel GenAI methods that validate PCBs from the security standpoint.

\subsubsection{Analog circuits}
PCBs contain a mix of both analog components, at the least to power the board, and digital components on it. Any trends or areas of concentrated GenAI work for analog circuits can translate to PCB contexts as well. 
Therefore, peering into automated analog circuit using GenAI can be useful for our analysis. Table~\ref{tab:relatedWorkcomparison} showcases 4 different analog circuit frameworks where GenAI models have proven useful.  

\textbf{Design: }
Out of the 4 identified works, 2 are analog circuit generation frameworks of which \cite{guo_circuit_2019} is a GAN-based topology generation model backed by knowledge graph representations of analog circuit. 
On the other hand, \cite{ho_denoising_2020} uses a diffusion model to diversify the limited available analog circuit images. While \cite{guo_circuit_2019} generates new functional analog circuit designs, \cite{ho_denoising_2020} augments the available dataset through a denoising method and creates new circuit images for detection frameworks to use.
These two works illustrate the differences in framework outputs depending on the type of task GenAI models are assigned to automate. 

\textbf{Optimisation: }
Both the design generation frameworks also incorporate some optimisation capabilities as \cite{guo_circuit_2019} incorporates enumeration methods in its knowledge graph to link between the various circuit components and its connections.
Meanwhile, \cite{ho_denoising_2020} uses the auto-regressive decoding method on top of its proposed probabilistic method to augment image-based analog circuits. 
Again, depending on the goal of optimisation, GenAI can either improve analog circuit generation or the quality of augmented samples.

\textbf{Validation: } 
3 out of the 4 collected analog circuit works perform functional validation. All of them are machine learning based algorithms, of which 2 are useful for analog circuit fault diagnosis. 
2 are also GAN-based.
One of the GAN-based frameworks \cite{guo_circuit_2019} generates analog circuit topologies and validates its generated design against design and layout rule checks, while the other GAN-based framework \cite{he_generative_2020} detects flaws from images of the inputted analog circuit. 

\textit{It can be inferred from these collected works that for analog circuit generation, GANs take precedence over LLMs.}
They offer more promising results, likely due to the image-based nature of circuit schematics. 
Furthermore, representing such circuits in LLM-generated code-based formats, especially for larger circuits, might not provide the complete circuit topology. 
None of these works are autonomous validation methods, and each approach requires human intervention. 
It can be best understood as a tool that assists verification engineers to make informed design decisions.  

\textbf{Security: }
\textit{None of the identified analog circuit works touch upon any security related concerns or constraints that the proposed GenAI works must navigate their way around.}
One possible explanation for this gap from the analog circuit works we have analysed is that GenAI has not shown enough success in autonomous functional validation itself. 
Security validation comes with an added layer of complexity, which is identifying and tackling the security vulnerabilities on a PCB, a task GenAI models cannot be entrusted with in their current states.

\begin{takeaway}
In embedded and cyber-physical systems, hardware design generation primarily involves LLMs. In analog circuits, more efforts are made to diversify the analog dataset using GANs, rather than on circuit design itself. For both domains, security relevant concerns are underexplored by GenAI applications. 
\end{takeaway}

\subsection{PCB design life cycle}
In the current EDA ecosystem, LLMs are the predominantly used GenAI model that generate code to implement electronic designs. 
However, from our taxonomy proposed in Section~\ref{sec:applying-taxonomy} for each of the PCB manufacturing stages, against the `Design' and `Optimisation' categories, we see that LLMs can be applied to a variety of PCB tasks. Some examples we have identified are high-level design generation \cite{jansen_words_2023}, in schematic and netlist generation \cite{liu_ladac_2024}, optimisation of the circuit topology \cite{vijayaraghavan_circuitsynth_2024}, fine-tuning the LLM to generate better circuit designs or improving LLM circuit generation performance \cite{raval_circuit_2025}~\cite{matsuo_schemato_2024}, and in automating and optimising layout and routing placement \cite{you_interactive_2024}~\cite{chen_pcbagent_2025}.

Our taxonomy also showcases various GenAI models and their applications in the PCB design life cycle, like the application of LLM-aided machine learning models for automating PCB layout placement \cite{chen_pcbagent_2025}, and GANs to generate new PCB samples in order to diversify the existing anomaly datasets \cite{huang_defect_2025}.
These observations correlate with the versatility of GenAI models, and support the argument that GenAI can be used in a variety of fields and domains, especially for classification and reasoning tasks. 

At the \textbf{design specification} stage of our taxonomy outlined in Table~\ref{tab:taxonomised_DesignSpec}, LLMs are identified as the biggest contributor to automated PCB design generation. 
Such an observation is expected, considering design specification files are generally text-based documents that can be fed to LLM prompts with ease. 

What is interesting to note is that at the \textbf{circuit schematic and/or netlist} generation stage, instead of noticing an upward trend in image-based GenAI models, LLMs continue to be on the rise. 
Some works like \cite{chang_lamagic_2024}, \cite{liu_ampagent_2024} and \cite{matsuo_schemato_2024} particularly focus on design generation using text-based representations of the circuit. 
This finding contrasts what we have observed in the analog circuit space, where GANs took precedence over LLMs. 
One possible explanation for this trend variation is that the LLMs used in these works are tasked with creating modular circuits. 

For smaller circuits, it is possible to represent all the contextual information of the circuit within a self-contained netlist that can be inputted into an LLM prompt. 
However, as the complexity of circuits increases, representing all the relevant circuit information in netlist format becomes a challenge, and increases the chances of LLMs to hallucinate or misunderstand parts of the inputted circuit \cite{sabree_openai_2024}. 
Integrating LLMs with ML models that have the ability to scan circuit images, like in \cite{bhandari_masala-chai_2025}, can help automate circuit design but the trade-off is the high computation power \cite{solorzano_environment-adaptable_2022} required to run ML-based LLM models, in comparison to lightweight LLMs.

In Table~\ref{tab:taxonomised_LayoutRouting}, we note that at the \textbf{layout and routing} stage, GenAI makes fewer attempts at design generation than the previous PCB stages. 
One reason for it is that PCBs inherently possess surface irregularity problems \cite{chen_pcbagent_2025}. 
These can vary in size and shape depending on the intended application of the PCB, the number of components on it, and their orientation \cite{liu_layoutcopilot_2025}. 
Since current literature does not enforce any strict layout protocols that all PCB manufacturers must follow either, it is harder for automated models to learn good circuit layout and routing practices \cite{li_fanoutnet_2023}.

\begin{takeaway}
As the complexity of the PCB design increases, GenAI model capabilities are limited by how circuits can be represented and their abilities to generate and optimise novel designs they have not been trained on. There is also not enough enforcement on what design rules are standard for PCBs. We believe further research in understanding how analysing and automating PCB design and optimisation stages could reveal further methods for GenAI assistance and even the potential for full automation. %
\end{takeaway}

\subsection{PCB test life cycle}
In the \textbf{design specification} stage, only 1 validation work is noted amongst the 5 collected ones. 
Even this validation framework \cite{liu_physics-informed_2024} uses an interactive chatbot to validate designs with the help of a validation engineer.

5 out of 16 works in the \textbf{circuit schematic and netlist} stage deal with functional validation. 
Of these, 3 are simulation-based inspection strategies, which are helpful in ensuring the circuit is a closed-loop and in validating expected circuit behaviour. 
The other 2 are LLM-powered SPICE tools, which do not support integration into existing EDA software yet \cite{abdollahi_hardware_2024}. 
All of these works require validation engineers to either intervene or double-check that outputted circuit behaviour is as expected, just like in the design specification stage.

For the \textbf{layout and routing} stage, 3 of the 5 works are tools that validation engineers can use to perform checks at the layout stage. 
Unfortunately, none can autonomously confirm that the resultant PCB at this stage is defect-free.
In their current ability to generate designs, these models require the expertise of a validation engineer in order to check them off before the resultant design passes to the next stage of PCB manufacturing. 

Only 1 work is outlined in the \textbf{printing} stage, which validates characters etched on the PCB. 
The ML based approach uses GANs to identify misprints on the board, but does not use LLMs, VLMs or any other language model whose prompts can be automated. 
Again, human validation is necessary to approve the identified flaw as indeed a defect. 

At the \textbf{assembly} stage, since manufacturing is complete, we identify all the works as validation based works. 
Out of the 8 listed works, 2 use diffusion models to classify defects, while the rest use GANs to create samples, or language models incorporated into machine learning algorithms or computer vision models to detect those anomalies. 
Therefore, all these strategies are more computationally intensive and expensive to run \cite{l_artificial_2025} than state of the art LLMs that process only text inputs.  

From our taxonomy tables in Section~\ref{sec:applying-taxonomy}, we see that a validation engineer can use GenAI tools to aid them in their judgment, but the final design decisions cannot be made by the GenAI models alone. 
Security risk assessments are also not explored until the \textbf{installation} stage of our taxonomy, when a functional hardware system is ready to be distributed. Here, GenAI works illustrate how hardware Trojans can be inserted \cite{kokolakis_harnessing_2024}, and how current validation methods of hardware circuit designs can be faulty \cite{gohil_llmpirate_2024}, highlighting gaps in the security validation space at specifically the PCB manufacturing stages. 

\begin{takeaway}
GenAI based work in PCB functional validation trends towards assistance with defect detection and quality control of the manufactured boards, but few works have aimed for fully-autonomous frameworks. Further, using GenAI to perform security validation at any of the manufacturing stages is an extremely underexplored area. This aligns with our identified gap on security validation with analog circuits and hardware systems as a whole. 
\end{takeaway}

In the next section, we will discuss the identified automation challenges in the context of the PCB design and test life cycle, and the current limitations of GenAI that hinder PCB manufacturing automation.

\section{Identified Challenges}
\label{sec:challenges-for-genai-usage}

This survey takes into consideration a total of 227 papers and analyses them against our proposed taxonomy in Section~\ref{sec:applying-taxonomy}. Of this, only 80 were of direct significance. In surveying these works, we noted some of the challenges that state-of-the-art GenAI models pose when applied to tasks in PCB Design and Test.
We also discuss broadly what the limitations of a narrow survey scope entail.

\subsection{Limited context windows for GenAI}
Current GenAI models -- especially LLMs -- are token-based \cite{rittikulsittichai_intelligent_2025}, which is a method of representation of the given inputs. In order for GenAI models to accurately represent or automate PCB designs, all the necessary PCB information must be converted into an interpretable format that can fit into the limited context windows of GenAI models \cite{gohil_llmpirate_2024}. Especially at the circuit schematic generation stage or in netlist to schematic automation operations, this poses concerns for more complex PCBs, as not all components and their associated nets can be concisely represented. 

\subsection{Insufficient domain-specific knowledge}
Despite GenAI's advancement in the hardware space, a lack of domain-specific knowledge hinders automation in the electronic design domain \cite{jin_wiseeda_2025} \cite{lidback_evaluation_2024}. This applies to PCBs as well, as PCBs contain a mix of analog (for eg: power sources, resistors, capacitors) and digital components (for eg: integrated circuits) in their design. Since state-of-the-art GenAI models are not necessarily trained on either of these datasets, the outputs generated by these models are results of hallucinations \cite{sabree_openai_2024}, and the results are often error-prone \cite{xiao_prefixllm_2024}. Some strategies like fine-tuning GenAI models on specific datasets or using few-shot prompting are some ways to improve model efficiency and its reasoning capabilities, as highlighted in previous sections. However, neither method can be solely relied on without a human expert to validate that the GenAI model prediction or generated output is correct.

Within the PCB design and test lifecycle, GenAI is currently unable to match the quality of outputs that hardware engineers can propose, and almost every stage of the life cycle demands human intervention in its validation work, as noted in each of the taxonomy stages.

\subsection{Inadequate GenAI integration with current PCB tools}
Most current PCB tools, like KiCAD and FreePCB, have been in the market for at least two decades, while GenAI has been a recently emergent field. Therefore, many of these tools do not easily allow GenAI models to integrate with them, even though efforts are being made to support this \cite{xiang_digital_2024}. A human is required in the pipeline to ensure that outputs from one tool can be fed into the other 
. Considering the current state of GenAI models, their limited context windows and a lack of domain-specific knowledge, these automation strategies powered by GenAI may be best viewed as a tool that reduces manual effort rather than replacing humans altogether.

\subsection{GenAI is missing security awareness}
We have already touched on how a lack of domain-specific knowledge for GenAI models decreases the legitimacy of its outputs. 
While efforts are being made to establish security rules and trustworthy guidelines for design, especially at the chip level \cite{wang_llms_2024}, the attack and defense perspectives for PCBs as a whole are still underexplored. Our insights from Section~\ref{sec:discussions:neighboringDomainTrends} aligns with this understanding as novel strategies to counter security threats have still not been automated using GenAI. 
What constitutes a security bug is also highly subjective to: (1) the resultant design and (2) the intended application the PCB is used for \cite{asadizanjani_physical_2021}. 
As a result, GenAI models do not have all the relevant context necessary to flag security vulnerabilities in design in their current state. 

Section~\ref{sec:RelatedWork:elecDesignAndTest} illustrates that automated security validation is a gap for hardware systems, which our taxonomy in Section~\ref{sec:applying-taxonomy} attests to in the PCB context as well. Therefore, in its current state, GenAI cannot completely replace human validation and verification in both the functionality and security domains of hardware design generation.

\subsection{(Survey Limitation) Narrow scope}
To our knowledge, there are no other surveys for PCBs and LLM applications on them, or GenAI applications on PCBs. Though lots of related work in the GenAI space, especially LLMs, apply to PCBs, no existing work classifies or taxonomises them into an application-independent repeatable categorization method like ours does. 

Our selection of 80 papers met all the selection criteria detailed in Section~\ref{sec:scope:selection}. Amongst the remaining analysed but excluded works, many PCB defect detection models included deep neural networks such as CNNs and RNNs which are not exactly GenAI. Considering any PCB fault detection frameworks, that are computer vision or ML based, without a focus on GenAI, would have exploded the scope of our research. There are also existing surveys in PCB defect detection that already taxonomise computer vision or machine-learning-based works, so such a contribution would not be novel.

In this section, we discussed how GenAI, while promising in many aspects, cannot independently automate all stages of the PCB design and test life cycle. However, the state-of-the-art models can be used as-is in conjunction with experienced engineers in stages where repetitive manual efforts can be replaced by an automated model. Such an integrated GenAI approach decreases the burden on human resources in the PCB design and test life cycle, and frees engineers up for tasks where their expertise is useful. 

\section{Future Research Directions and Opportunities}
\label{sec:future-directions}

\subsection{Taxonomising GenAI-based IC Design and Test}
In Section~\ref{sec:Background} and Table~\ref{tab:llm_eda_survey}, we analysed that current automation trends in the EDA domain are geared towards integrated circuit design.
Predominantly, we found that LLMs are the most used GenAI models in this space, because of their ability to generate code like HDL which can describe and implement chip designs.
However, board related automation and other electronic based automation also falls under EDA, although they are not tagged as EDA contributions in literature.

Our prescribed taxonomy in Section~\ref{sec:applying-taxonomy} showcases how there are other aspects to electronic design as well, that are not limited to code-based IC representations.
EDA is hence an area that can benefit not only from more contributions to automating other electronic system design processes, but also in applying other GenAI models to specialised aspects of design.
In turn, our application-independent taxonomy can be applied to those works, classifying them in a manner similar to the PCB design and test life cycle. 

\subsection{Lack of security oriented validation approaches in current GenAI works} 
We have discussed how security validation is an underresearched area, both in the context of analog circuit design in Section~\ref{sec:Background} and within the PCB design and test life cycle. 
No surveyed works in Section~\ref{sec:applying-taxonomy} targeted both functional and security validation. 
In other words, the taxonomised works that perform validation either contribute to functional testing, or security validation, but never both. 
One possible explanation for this gap is the differing requirements between ensuring that a system meets its functional specifications (e.g. I/O, timing, application logic), and determining whether any components on the board could be a vector for malicious misuse or tampering.
Added to this is the limited amount of domain-specific knowledge, as described in Section~\ref{sec:challenges-for-genai-usage}, that hinder training and learning capabilities for the GenAI models.
Contributions to this space will benefit PCB manufacturing automation and reduce the number of PCB attacks and hardware Trojans that make their way through the design life cycle.

\section{Conclusion \label{sec:Conclusion}}

This research explores the various applications of GenAI models in the current PCB design and test life cycle. We proposed a taxonomy to classify these collected works based on their goal: to design, to optimise or to validate PCBs in the supply chain in one of two ways: (1) functionally, or (2) from the security point of view. 

RQ1 is answered in Section~\ref{sec:applying-taxonomy}, which highlights that most current works applying GenAI to PCB operations focus on design and optimisation. 
There are fewer works that apply GenAI to the necessary area of validation. 
This may be based on the amounts of data required and available for GenAI training, as well as the tendencies to hallucinate or generate error-prone outputs. 

Through classifying the 227 academic and industry works, of which 80 are included in this survey, we outlined the current obstacles and challenges at the intersection of GenAI applied to the PCB design and test life cycle in Section~\ref{sec:challenges-for-genai-usage}. We highlighted several major limitations: (1) how limited context windows on text-based GenAI models restrict how complex circuits can be represented and inputted, (2) how insufficient domain data makes training GenAI models challenging, and (3) how current GenAI models lack the security awareness needed to detect security-based vulnerabilities in design. 

By classifying current research into a taxonomy proposed in Section~\ref{sec:applying-taxonomy}, we determined research opportunities and directions for the application of PCB design tasks. Current research opportunities exist in improving security validation methods using GenAI, extending GenAI models to other parts of the EDA domain such as electronic or hardware design rather than chip-focused works, as well as applying other models instead of just LLMs to the EDA space. 

We believe that our taxonomy need not be limited solely to PCB design and test, though this was our primary focus. Given the overlap between this area and other hardware life cycles (e.g. for integrated circuits), we believe a similar study could be undertaken to explore these and parallel areas.

\bibliographystyle{ACM-Reference-Format}
\bibliography{refs/pcb_elec_sats}

\end{document}